\documentclass[smallextended]{svjour3}

\usepackage[misc]{ifsym}
\usepackage{natbib}
\usepackage[T1]{fontenc}
\usepackage[bookmarks=false,colorlinks,linkcolor=blue,citecolor=blue,urlcolor=blue]{hyperref}
\usepackage{url}
\usepackage{xspace}
\usepackage{numprint}
\npthousandsep{,}

\AtBeginDocument{%
}

\usepackage{booktabs}
\usepackage{multirow}
\usepackage{ltablex}
\usepackage{threeparttable}
\usepackage{footnotehyper}
\makesavenoteenv{table}
\makesavenoteenv{longtable}
\newcolumntype{R}[1]{>{\raggedleft\arraybackslash}p{#1}}
\newcolumntype{L}[1]{>{\raggedright\arraybackslash}p{#1}}

\usepackage{graphicx}
\usepackage{subcaption}
\usepackage[export]{adjustbox}

\newcommand{\figbox}[1]{\begin{tcolorbox}
[breakable,enhanced,colback=white,arc=0pt,boxrule=0.5pt,boxsep=4pt,left=0pt,right=0pt,top=0pt,bottom=0pt,after={\vspace{-0.1cm}}]
{#1}\end{tcolorbox}}

\usepackage{tcolorbox,tikz}
\usetikzlibrary{patterns}
\usepackage{pgfplotstable}

\usepackage{amssymb,amsfonts}
\usepackage{algorithm}
\usepackage{algorithmicx}
\usepackage{algpseudocode}
\usepackage{listings}
\usepackage{color}
\definecolor{bluekeywords}{rgb}{0.13, 0.13, 1}
\definecolor{greencomments}{rgb}{0, 0.5, 0}
\definecolor{redstrings}{rgb}{0.9, 0, 0}
\definecolor{graynumbers}{rgb}{0.5, 0.5, 0.5}
\newcommand{\answer}[1]{\begin{tcolorbox}
[enhanced,colback=white,boxrule=0.5pt,boxsep=4pt,left=0pt,right=0pt,top=0pt,bottom=0pt,after={\vspace{-0.1cm}}]
{#1}\end{tcolorbox}}

\tcbuselibrary{skins}
\tcbuselibrary{breakable}

\usepackage{fp}
\newcommand{\Mul}[2]{
  \FPeval{vTemp}{#1*#2}%
  \FPround\vTemp{\vTemp}{0}%
  \ifnum \vTemp<10%
      {\hspace*{4.3pt}\FPprint{vTemp}}%
  \else%
      {\FPprint{vTemp}}%
  \fi%
}%

\newcommand{\ShowAbsoluteNumber}[1]{%
    \ifnum #1<10%
        {\hspace*{14.6pt}#1}%
    \else%
        \ifnum #1<100%
            {\hspace*{11pt}#1}%
        \else%
            \ifnum #1<1000%
                {\hspace*{7pt}#1}%
            \else%
                {\numprint{#1}}%
            \fi%
        \fi%
    \fi%
}%

\newcommand{\ShowPercentage}[2]{%
    \ifnum #1=0%
        {~(\hspace*{6pt}\FPprint{0}\%)}%
    \else%
        \FPeval\percentage{round(#1/#2*100,0)}%
        \ifnum \percentage=0%
            {(\hspace*{2pt}<1\%)}%
        \else%
            \ifnum \percentage<10%
                {(\hspace*{6pt}\FPprint{percentage}\%)}%
            \else%
                \ifnum \percentage<100%
                    {(\hspace*{4pt}\FPprint{percentage}\%)}%
                \else%
                    {(\FPprint{100}\%)}%
                \fi%
            \fi%
        \fi%
    \fi%
}%

\newcommand{\ShowPercentageNoHundred}[2]{%
    \ifnum #1=0%
        {(\hspace*{4.3pt}\FPprint{0}\%)}%
    \else%
        \FPeval\percentage{round(#1/#2*100,0)}%
        \ifnum \percentage=0%
            {(\mbox{\scalebox{.65}{$<$}}1\%)}%
        \else%
            \ifnum \percentage<10%
                {(\hspace*{4.3pt}\FPprint{percentage}\%)}%
            \else%
                {(\FPprint{percentage}\%)}%
            \fi%
        \fi%
    \fi%
}%

\newcommand{\ShowPercentageNoHundredNoParentheses}[2]{%
    \ifnum #1=0%
        {\hspace*{5.2pt}\FPprint{0}\%}%
    \else%
        \FPeval\percentage{round(#1/#2*100,0)}%
        \ifnum \percentage=0%
            {\hspace*{-1.3pt}<1\%}%
        \else%
            \ifnum \percentage<10%
                {\hspace*{5.2pt}\FPprint{percentage}\%}%
            \else%
                {\hspace*{1pt}\FPprint{percentage}\%}%
            \fi%
        \fi%
    \fi%
}%

\usepackage{xcolor}
\newlength\MAX  
\newcommand*\Chart[1]{\Mul{#1}{100}\%~\rlap{\textcolor{black!20}{\rule{\MAX}{2ex}}}\rule{#1\MAX}{2ex}}

\newcommand{\inlinechart}[2]{%
\FPeval{\BLACKBARSIZE}{#1/#2}\textcolor{black!80}{\rule{\BLACKBARSIZE\BARSIZE}{1.6ex}}%
\FPeval{\BLACKBARSIZE}{1 - (#1/#2)}\textcolor{black!10}{\rule{\BLACKBARSIZE\BARSIZE}{1.6ex}}%
}

\newlength\BARSIZE  
\newcommand*\ChartWithAbsoluteAndPercentageNumbers[2]{
    \numprint{#1} & \ShowPercentageNoHundred{#1}{#2} & \inlinechart{#1}{#2}
}

\newlength\BARSIZESIMPLE  
\newcommand*\ChartWithPercentageOnly[2]{
    \ShowPercentageNoHundredNoParentheses{#1}{#2}
    \rlap{\textcolor{black!20}{\rule{\BARSIZESIMPLE}{2ex}}}\FPeval{\BLACKBARSIZE}{#1/#2}\rule{\BLACKBARSIZE\BARSIZESIMPLE}{2ex}
}

\usepackage[]{xcolor}
\newcommand{\TODO}[1]{\textcolor{red}{#1}\GenericWarning{}{LaTeX Warning: TODO: #1}}\newcommand\todo\TODO

\newcommand{\Styler}{\textsc{Styler}\xspace}
\newcommand{\Naturalize}{\textsc{Naturalize}\xspace}
\newcommand{\CodeBuff}{\textsc{CodeBuff}\xspace}
\newcommand{\CheckStyleIDEA}{\textsc{CheckStyle-IDEA}\xspace}
\newcommand{\CSIDEA}{\textsc{CS-IDEA}\xspace}

\newcommand{\checkstylexml}{\texttt{checkstyle.xml}\xspace}

\newcommand{\protocolOne}{$Styler_{random}$\xspace}
\newcommand{\protocolTwo}{$Styler_{3grams}$\xspace}

\newcommand{\nbProjectsTotalToPrint}{\numprint{105}\xspace}
\newcommand{\nbViolationsTotalToPrint}{\numprint{27058}\xspace}

\newcommand{\nbProjectsEvaluation}{104} % without the project for calibration
\newcommand{\nbViolationsEvaluation}{26791} % without the ones for calibration
\newcommand{\nbProjectsEvaluationToPrint}{\numprint{\nbProjectsEvaluation}\xspace}
\newcommand{\nbViolationsEvaluationToPrint}{\numprint{\nbViolationsEvaluation}\xspace}

\newcommand{\nbViolationsCalibration}{267} % without the ones for evaluation
\newcommand{\nbViolationsCalibrationToPrint}{\numprint{\nbViolationsCalibration}\xspace}

\begin{document}

\title{Styler: learning formatting conventions to repair Checkstyle violations}

\author{Benjamin Loriot \and Fernanda Madeiral \and Martin Monperrus}

\institute{Benjamin Loriot \at
            University of Technology of Compi\`egne\\ Compi\`egne, France \\
            \email{bloriot97@gmail.com}
            \and
            \Letter~Fernanda Madeiral \at
            KTH Royal Institute of Technology\\ Stockholm, Sweden \\
            \email{fer.madeiral@gmail.com}
            \and
            Martin Monperrus \at
            KTH Royal Institute of Technology\\ Stockholm, Sweden \\
            \email{martin.monperrus@csc.kth.se}
}

%\date{Received: date / Accepted: date}

\date{Empirical Software Engineering, 2022. DOI: 10.1007/s10664-021-10107-0}

\maketitle

\begin{abstract}
Ensuring the consistent usage of formatting conventions is an important aspect of modern software quality assurance.
To do so, the source code of a project should be checked against the formatting conventions (or rules) adopted by its development team, and then the detected violations should be repaired if any. While the former task can be automatically done by format checkers implemented in linters, there is no satisfactory solution for the latter. Manually fixing formatting convention violations is a waste of developer time and code formatters do not take into account the conventions adopted and configured by developers for the used linter.
In this paper, we present \Styler, a tool dedicated to fixing formatting rule violations raised by format checkers using a machine learning approach. For a given project, \Styler first generates training data by injecting violations of the project-specific rules in violation-free source code files. Then, it learns fixes by feeding long short-term memory neural networks with the training data encoded into token sequences. Finally, it predicts fixes for real formatting violations with the trained models. Currently, \Styler supports a single checker, Checkstyle, which is a highly configurable and popular format checker for Java. In an empirical evaluation, \Styler repaired 41\% of \nbViolationsEvaluationToPrint Checkstyle violations mined from \nbProjectsEvaluationToPrint GitHub projects. Moreover, we compared \Styler with the IntelliJ plugin \CheckStyleIDEA and the machine-learning-based code formatters \Naturalize and \CodeBuff. We found out that \Styler fixes violations of a diverse set of Checkstyle rules (24/25 rules), generates smaller repairs in comparison to the other systems, and predicts repairs in seconds once trained on a project. Through a manual analysis, we identified cases in which \Styler does not succeed to generate correct repairs, which can guide further improvements in \Styler. Finally, the results suggest that \Styler can be useful to help developers repair Checkstyle formatting violations.
\end{abstract}

\section{Introduction}

% Coding conventions' importance
Coding conventions are widely recognized as a means to improve the internal quality of software systems \citep{Prause2015}.
They are rules that developers agree on for writing code, which encode best coding practices, widely adopted standards, or developers' preferences. The usage of coding conventions helps to reduce style deviations, which are nothing but distracting noise when reading code \citep{Spinellis2011,Prause2015}.

% How to keep the code consistent: violation detection and violation repair
However, keeping all source code files of a project compliant with the coding conventions agreed by a development team is a challenge.
For that, two main activities must be performed: the \textit{detection}
and the \textit{repair} of coding convention \textit{violations}.
The detection of coding convention violations can be automatically performed using \textit{linters}. A linter is a tool that statically analyzes code to check its compliance with rules and warns software developers when rule violations are found. The usage of linters also brings challenges because the developers need to create a configuration according to their adopted conventions so that the linter detects the right violations (not more and not less). Nevertheless, in this paper, we focus on the latter task, i.e., the repair of violations, which is a little researched problem.

% The problem of the two existing ways to repair linter violation
To repair coding convention violations, developers can either perform fixes manually or use automated solutions that produce fixes. Manually fixing these violations is a waste of valuable developer time. Considering \textit{formatting convention violations}, which are the focus of this paper, developers could use \textit{code formatters} as automated solution. However, this alternative is also not satisfactory. With code formatters, the key problem is that they do not take into account the project-specific rules, those that are configured by developers for the used linter.

% Problem statement
Inspired by the problem statement of program repair \citep{Monperrus2018}, we state in this paper the problem of automatically repairing formatting violations: \textit{given a program, a set of format checker rules, and one rule violation, the goal is to modify the source code formatting so that no violation is raised by the format checker}. A \textit{format checker} is a linter, or a part of a linter, that focuses on formatting checks, since linters cover several classes of coding conventions, e.g., naming and formatting.

% Styler
In this paper, we explore that problem in the context of Checkstyle\footnote{\url{https://checkstyle.sourceforge.io/}, last access: 2020-07-13}, a popular format checker for the Java language.
We present \Styler, a tool dedicated to fixing formatting violations in Java source code. The uniqueness of \Styler is its applicability to any formatting convention because its approach is not based on specific format checker rules. The key idea behind \Styler is the usage of machine learning to learn the formatting conventions that are used in a software project. The learning is based on training data generated by \Styler through the modification of source code files to trigger violations of the formatting rules configured by developers for a given project. Once trained, \Styler predicts changes on formatting characters (e.g., whitespaces) to fix formatting convention violations happening in the wild.
Technically, \Styler encodes Java source code containing formatting violations into abstract token sequences and uses sequence-to-sequence machine learning models based on long short-term memory neural networks (LSTMs).

% Experiment and findings
To evaluate \Styler, we conducted a large scale experiment using \nbViolationsEvaluationToPrint Checkstyle formatting violations mined from \nbProjectsEvaluationToPrint GitHub projects.
Based on our research questions, we found out that \Styler repairs many violations (41\%) from a diverse set of formatting rules (24/25). It generally performs better for fixing violations related to horizontal whitespace between Java tokens than violations related to tabulations and line length. Moreover, \Styler produces smaller repairs compared to the state-of-the-art machine learning formatters \citep{Allamanis2014naturalize,Parr2016codebuff} and the IntelliJ plugin \CheckStyleIDEA \citep{checkstyle-idea}. Finally, \Styler repairs violations in seconds, once it is trained for a given project.

% Summary of the contributions
To sum up, our contributions are:

\begin{itemize}
\item A novel approach to fix violations of code formatting conventions, based on machine learning. The approach is able to learn project-specific formatting rules with a self-training data generation strategy and repair formatting rule violations with a sequence-to-sequence machine learning model;

\item A tool, called \Styler, which implements our approach in the context of Java and Checkstyle, to repair Checkstyle formatting violations. The tool is made publicly available\footnote{\url{https://github.com/KTH/styler/}} for future research and usage;

\item A dataset of real-world Checkstyle violations mined from GitHub repositories. The dataset is publicly available\footnote{\url{https://github.com/KTH/checkstylerr/}} for future research;

\item A comparative experiment of the performance of \Styler against the state-of-the-art code formatters \citep{checkstyle-idea,Allamanis2014naturalize,Parr2016codebuff}. The results of the experiment are also publicly available\footnote{\url{https://kth.github.io/styler-experiments/}} for the sake of open science.
\end{itemize}

\newpage

% Paper outline
The remainder of this paper is organized as follows. \autoref{sec:background} presents the background of this work. \autoref{sec:styler} presents \Styler in detail, including its workflow and technical principles. \autoref{sec:evaluation-design} presents the design of our experiment for evaluating \Styler and comparing it with three code formatters. The experimental results are then presented in \autoref{sec:evaluation-results}. \autoref{sec:discussion} presents discussions, and \autoref{sec:related-work} presents the related works. Finally, \autoref{sec:conclusion} presents the concluding remarks of this work.

\section{Background}\label{sec:background}

Coding conventions play an important role in software development and maintenance. In this section, we present a background on coding conventions and tools that help developers enforce them. In addition, we report on a study of the usage of Checkstyle, a tool that statically checks Java code against a specified set of coding conventions.

\subsection{Coding conventions}\label{sec:background:coding-conventions}

\textit{Coding conventions}, also known as coding style and coding standards, are \textit{rules} that developers agree on for writing code. The usage of coding conventions does not affect the behavior of software systems. Instead, developers use them to improve code readability and maintainability. Although not all coding practices are perceived by developers as enhancing code readability \citep{SantosAndGerosa2018}, they help to reduce style deviations, which are nothing but distracting noise when reading code \citep{Spinellis2011,Prause2015}.

There are several kinds of coding conventions, e.g., conventions related to naming and formatting. In this paper, we focus on the latter, i.e., \textit{formatting} conventions. Formatting refers to the appearance or the presentation of the source code. One can change the formatting by using, for instance, \textit{non-printable characters}, such as spaces, tabulations, and line breaks.
In free-format languages such as Java and C++, the code formatting does not change the abstract syntax tree of programs. In non-free-format languages, such as Python and Haskell, the formatting is even related to behavior, which means that correcting formatting issues can fix bugs.

To exemplify formatting conventions, consider \autoref{fig:preferences}, which shows two well-known ways that developers may follow when placing left curly braces in code blocks. Note that one way is to place the left curly brace in a new line (\autoref{fig:preferences-a}) while another way is to place it at the end of the conditional expression line (\autoref{fig:preferences-b}). The way to actually do it in a software project depends on what the project's development team chooses. Agreeing on coding conventions to be followed in a software project is important to avoid edit wars and endless debates.

\begin{figure}[t]
    \centering
    \begin{subfigure}[b]{.45\linewidth}
        \centering
        \begin{lstlisting}[xleftmargin=.2\textwidth]
if (condition)
{
    // do something
}
        \end{lstlisting}
        \caption{Left curly in a new line.}
        \label{fig:preferences-a}
    \end{subfigure}
    \begin{subfigure}[b]{.45\linewidth}
        \centering
        \begin{lstlisting}[xleftmargin=.2\textwidth]
if (condition) {
    // do something
}
        \end{lstlisting}
        \caption{Left curly at the end of line.}
        \label{fig:preferences-b}
    \end{subfigure}
    \caption{Two conventions for placing a left curly brace.}
    \label{fig:preferences}
\end{figure}

\subsection{Detection of coding convention violations}

A challenge faced by developers is to keep their code compliant with the agreed coding conventions. Basically, every new change in the code must satisfy the adopted coding conventions.
Manual analysis of code changes for checking if they do not violate the adopted coding conventions is time-consuming and error-prone.
To do so automatically, one can use \textit{linters}.
A linter is a tool that statically analyzes code to check its compliance with rules and warns software developers when rule violations are found. The rules might be related to functional problems, such as resource leakage or incorrect logic, and maintainability problems, such as non-compliance with best practices or violations of style conventions \citep{Beller2016}. As a side note, the literature does not consistently relate linters and automated static analysis tools (abbreviated as ASATs, also known as static analyzers). However, we understand that an automated static analysis tool is any tool that analyzes source code without the need to run it, including, for instance, tools for software analytics. Therefore, in this paper, we consider that any linter belongs to the family of automated static analysis tools, but that automated static analysis tools are not all about analyzing code against a set of rules.

Linters can be usually integrated into IDEs and build tools.
On the one hand, when integrated into IDEs, developers may manually run the linter before they commit their changes.
If they do not do it, they might face a lot of violations raised by the linter after the end of the building step for a release or for shipping the program.
On the other hand, when a linter is integrated into build tools, it might be automatically executed in Continuous Integration (CI) environments.
The important coding conventions might be configured to make CI builds break when they are violated. 
This way, developers are forced to repair coding convention violations early in the software development process.

Several linters have been developed for different programming languages. Some examples include
ESLint\footnote{\url{https://eslint.org/}, last access: 2020-07-13} for JavaScript,
Pylint\footnote{\url{https://www.pylint.org/}, last access: 2020-07-13} for Python,
StyleCop\footnote{\url{https://github.com/StyleCop/StyleCop/}, last access: 2020-07-13} for C\#, and
RuboCop\footnote{\url{https://docs.rubocop.org}, last access: 2020-07-13} for Ruby.
For Java, which is our target language in this paper, a commonly used linter is Checkstyle\footnote{\url{https://checkstyle.sourceforge.io/}, last access: 2020-07-13}. Checkstyle is composed of several \textit{checks} that encode style-related rules. For instance, Checkstyle contains a check named \emph{LeftCurly} that checks for the placement of left curly braces for code blocks, which is about the example illustrated in \autoref{fig:preferences}. Developers specify in a configuration file, then, their coding conventions by selecting and configuring Checkstyle checks. We refer to this configuration file as \textit{Checkstyle ruleset} and, hereafter, we refer to Checkstyle checks as Checkstyle rules.
Finally, Checkstyle is a flexible linter that can be integrated into IDEs (e.g., IntelliJ, Eclipse, and NetBeans) and build tools (e.g., Maven and Gradle).

\subsection{Usage of Checkstyle in the wild}\label{sec:study-checkstyle-usage}

Linters have been the subject of investigation in recent research \citep{Zampetti2017,Vassallo2018,Marcilio2019}. However, the existing studies did not investigate at scale and look into how style checking tools are specifically used. In this section, we present a study focused on the usage of Checkstyle, which is a popular linter for Java that checks source code style.

\subsubsection{Checkstyle usage in open-source projects}

\noindent\textit{Method.}
To measure the usage of Checkstyle, we queried GitHub\footnote{On June 23, 2021.} to only retrieve Java projects with at least five stars, because stars have been shown meaningful to sample projects from GitHub \citep{Beller2017}. We found \numprint{171195} Java projects. Then, we searched each of them\footnote{On June 23--24, 2021.} for finding a Checkstyle ruleset file.
Note that a Checkstyle ruleset file can have any name, but we followed a conservative approach towards identifying true positive files by using a set of commonly used names\footnote{Commonly used names for Checkstyle ruleset files: `checkstyle.xml', `.checkstyle.xml', `checkstyle\_rules.xml', `checkstyle\_config.xml', `checkstyle\_configuration.xml', `checkstyle\_checker.xml', `checkstyle\_checks.xml', `google\_checks.xml', `sun\_checks.xml'. Variants by replacing `\_' by `-' were also used.}. For simplicity, hereafter, we refer to a Checkstyle ruleset file as \checkstylexml.

\vspace{5pt}
\noindent\textit{Results.}
We found \numprint{4334} Java projects containing a \checkstylexml file, which is 2.53\% of all Java projects with at least five stars on GitHub.
\autoref{tab:build-tool-ci-usage} shows the proportion of these projects that use Maven, Gradle, or Ant as their build tools, and the Travis or Circle CI services. We note that build tools are widely used among projects using Checkstyle: 98\% of the projects use at least one build tool. Moreover, 44\% of the projects use a continuous integration service, which shows the software engineering maturity of the sampled projects.

\begin{table}[ht]
    \caption{Usage of build tools and CI services in the \numprint{4334} projects using Checkstyle.}
    \label{tab:build-tool-ci-usage}
    \centering
    \begin{tabular}{@{}l l p{.28\textwidth}@{}}
        \toprule
        \multirow{3}{*}{Build tool} & Maven & \Chart{0.54} \\
        & Gradle & \Chart{0.47} \\
        & Ant & \Chart{0.09} \\
        \midrule
        \multirow{2}{*}{CI service} & Travis CI & \Chart{0.41} \\
        & Circle CI & \Chart{0.04} \\
        \bottomrule
    \end{tabular}
\end{table}

\subsubsection{Popularity of Checkstyle rules}

\noindent\textit{Method.}
To check the usage of the 182 Checkstyle rules\footnote{The set of Checkstyle rules considered in our study is from the Checkstyle version 8.43 (released on May 30, 2021).}, we analyzed the previously-found \checkstylexml files from the \numprint{4334} projects.
Our goal is to investigate the most used rules and check if formatting-related rules, which are the target of this work, are widely used.

\vspace{5pt}
\noindent\textit{Results.}
We found out that all Checkstyle rules are used. \autoref{fig:top_10_rules_most_used} shows the top-10 most used rules. The bars in dark gray represent formatting-related rules, and the bars in light gray represent non-formatting rules. In addition, the bar in gray with a dot pattern represents a rule that can be about formatting, but it depends on how it is configured since it is a regex rule. In the top-10 most used rules, there are three rules related to formatting and one that can be. Notably, the top-3 most used rules are or can be formatting-related ones. Therefore, we conclude that formatting-related rules are important for developers, which validates the relevance of our work.

\pgfplotstableread[col sep=comma, header=true]{rules_usage.csv}{\datatable}

\pgfplotstablegetrowsof{\datatable}
\edef\numberofrows{\pgfplotsretval}

\pgfplotsset{%
    discard if not/.style 2 args={
        x filter/.code={
            \edef\tempa{\thisrow{#1}}
            \edef\tempb{#2}
            \ifx\tempa\tempb
            \else
                \def\pgfmathresult{inf}
            \fi
        }
    }
}

\begin{figure}[h!]
  \centering
  \begin{tikzpicture}
    \begin{axis}
    [xbar,
    width=0.89\linewidth,
    height=5.5cm,
    bar width=6.5pt,
    visualization depends on={x \as \originalvalue},
    point meta={x/4334*100},
    nodes near coords={\pgfmathprintnumber{\originalvalue}~~(\pgfmathprintnumber{\pgfplotspointmeta}\%)},
    nodes near coords align={horizontal},
	xlabel=\# Projects on GitHub,
    xlabel near ticks,
    xmax=4800,
    ytick={0,1,...,\numberofrows},
    yticklabels from table={\datatable}{rule},
    yticklabel style={align=right, text width=2.2cm},
    y dir=reverse,
    enlarge y limits=0.08,
    xtick pos=left,
    ytick pos=left,
    legend pos=south east,
    legend cell align={left},
    legend image code/.code={
        \draw [#1] (0cm,-0.1cm) rectangle (0.3cm,0.07cm);
    }]
    ]
    %\addplot[draw=black, fill=black!15] table[x=frequency, y expr=\coordindex]{\datatable};
    \addplot[bar shift=0pt, draw=black, fill=gray!100] table[discard if not={keyword}{Y}, x=frequency, y expr=\coordindex, col sep=comma]{rules_usage.csv};
    
    \addplot[bar shift=0pt, draw=black, fill=gray!30] table[discard if not={keyword}{N}, x=frequency, y expr=\coordindex, col sep=comma]{rules_usage.csv};
    
    \addplot[bar shift=0pt, draw=black, fill=gray!60, postaction={pattern=dots}] table[discard if not={keyword}{B}, x=frequency, y expr=\coordindex, col sep=comma]{rules_usage.csv};
    
    \legend{Formatting rules, Non-formatting rules, Both}
    \end{axis}
  \end{tikzpicture}
  \caption{The top-10 most used Checkstyle rules.}
  \label{fig:top_10_rules_most_used}
\end{figure}

\section{\Styler}\label{sec:styler}

\Styler is a tool dedicated to helping developers keep their source code compliant with their adopted formatting conventions by automatically fixing formatting violations in Java source code. \Styler could be used in different software development workflows. For instance, \Styler could be used locally as a pre-hook commit when developers are about to release projects. It could also be used in continuous integration environments, where pull requests could be automatically opened with formatting fixes' suggestions.
In this section, we present the workflow and the technical principles of \Styler.

\subsection{Targeted violation types}

\Styler is about learning and repairing violations related to \textit{formatting conventions}.
For instance, consider that a developer specified that left curly braces must always be placed at the end of lines (as shown in \autoref{fig:preferences-b}). If this rule is not satisfied (e.g., such as in \autoref{fig:preferences-a}), a given linter triggers a formatting-related violation: for instance, Checkstyle would output the violation presented in \autoref{fig:checkstyle-violation}, and SonarJava\footnote{\url{https://github.com/SonarSource/sonar-java}, last access: 2020-07-13} would find a violation of the rule ``An open curly brace should be located at the end of a line''\footnote{\url{https://rules.sonarsource.com/java/RSPEC-1105}, last access: 2020-07-13}.
In order to fix this violation, the line break before the token ``\{'' must be replaced by a single space.

As mentioned in \autoref{sec:background:coding-conventions}, there are different classes of conventions, e.g., formatting and naming, and consequently different automated checks in linters. In \Styler, we exclusively focus on \textit{formatting checks related to non-printable characters}, such as indentation and whitespace before and after punctuation. Hereafter, we refer to the linter part related to these formatting checks as \textit{format checker}.

\subsection{\Styler workflow}

\begin{figure}[ht]
    \centering
    \includegraphics[scale=0.7]{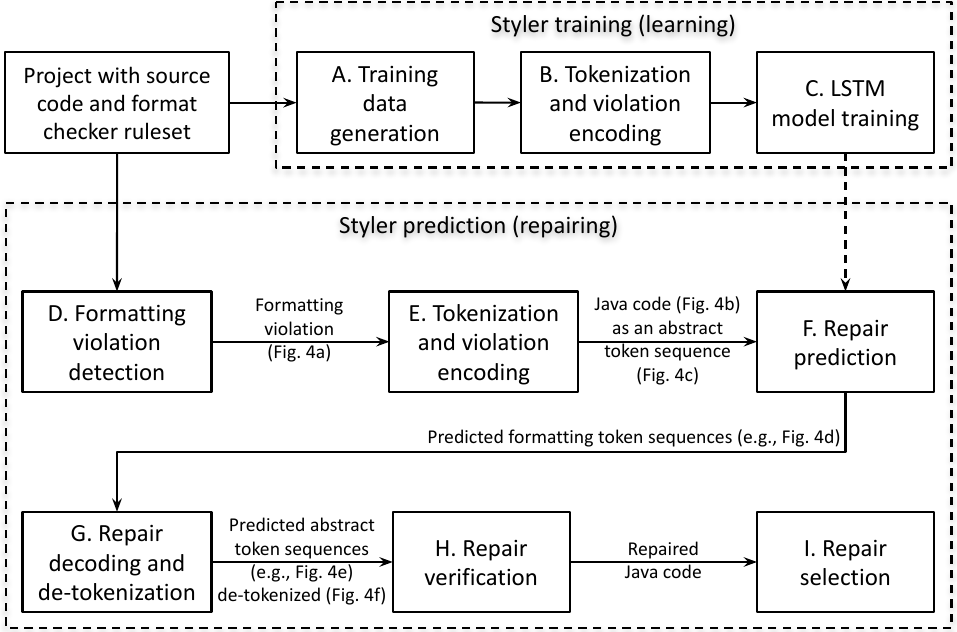}
    \caption{\Styler workflow.}
    \label{fig:styler}
\end{figure}

\autoref{fig:styler} shows the workflow of \Styler. It is composed of two main components: `\Styler training' for learning how to repair formatting violations and `\Styler prediction' for repairing a real formatting violation raised by a format checker. \Styler receives as input a software project, including its source code and its format checker ruleset.

The component `\Styler training' is responsible for learning how to repair formatting violations on the given project according to the project-specific format checker ruleset. It creates the training data by injecting formatting violations into violation-free source code files belonging to the project (step A).
Then, it encodes the training data into abstract token sequences (step B) in order to train LSTM neural networks (step C). The learned LSTM models are later used to predict repairs.

The component `\Styler prediction' is responsible for predicting fixes for real formatting violations. It first detects formatting violations by running the format checker on the project (step D).
Then, \Styler encodes the violating code into an abstract token sequence (step E), which is given as input to the LSTM models (step F) previously learned.
The models predict fixes for the given formatting violation. These fixes are in the format of formatting token sequences, so they are translated back to Java code (step G).
\Styler then runs the format checker on the new Java code containing the predicted fixes (step H). Finally, among the predicted fixes where no violation is raised by the format checker, \Styler selects one formatting fix to give as output (step I).
As \Styler only impacts the formatting of source code, its repairs do not change the behavior of the program under consideration.

\subsection{\Styler in action}

Consider the formatting violation presented in \autoref{fig:checkstyle-violation}.
This violation is about the Checkstyle \emph{LeftCurly} rule, which was configured to enforce that left curly braces are placed at the end of lines.
The Java source code that caused such a violation is presented in \autoref{subfig:violating-code}.

% ibinti-bugvm/0
\begin{figure}[t]
    \centering
    \begin{subfigure}[b]{1\linewidth}
        \figbox{
        \texttt{[ERROR] DSAEncoder.java:7:1: '\{' at column 1 should be on the previous line. [LeftCurly]}
        }
        \caption{Checkstyle \emph{LeftCurly} rule violation.}
        \label{fig:checkstyle-violation}
    \end{subfigure}
    \begin{subfigure}[b]{1\linewidth}
        \vspace{2pt}
        \begin{tcolorbox}[breakable,enhanced,colback=white,arc=0pt,boxrule=0.5pt,boxsep=0pt,left=0pt,right=0pt,top=-3pt,bottom=-3pt,after={\vspace{-0.1cm}}]
        \begin{lstlisting}[label={code:violating-code},xleftmargin=-28pt,basicstyle=\scriptsize]
       6 public interface DSAEncoder
       7 {
       8     byte[] encode(BigInteger r, BigInteger s)
        \end{lstlisting}
        \end{tcolorbox}
        \caption{Source code snippet with the violation.}
        \label{subfig:violating-code}
    \end{subfigure}
    \begin{subfigure}[b]{1\linewidth}
        \vspace{2pt}
        \figbox{
        \texttt{\textit{before-context-tokens} <LeftCurly> Identifier 1\_NL \{ 1\_NL\_4\_ID\_SP </LeftCurly> \textit{after-context-tokens}}
        }
        \caption{Original abstract token sequence.}
        \label{subfig:violating-code-tokenized}
    \end{subfigure}
    \begin{subfigure}[b]{1\linewidth}
        \vspace{2pt}
        \figbox{
        \texttt{1\_SP 1\_NL\_4\_ID\_SP}
        }
        \caption{Formatting token sequence generated by a LSTM model.}
        \label{subfig:patched-machine-learning-output}
    \end{subfigure}
    \begin{subfigure}[b]{1\linewidth}
        \vspace{2pt}
        \figbox{
        \texttt{\textit{before-context-tokens} <LeftCurly> Identifier 1\_SP \{ 1\_NL\_4\_ID\_SP </LeftCurly> \textit{after-context-tokens}}
        }
        \caption{Predicted abstract token sequence.}
        \label{subfig:patched-code-tokenized}
    \end{subfigure}
    \begin{subfigure}[b]{1\linewidth}
        \vspace{2pt}
        \begin{tcolorbox}[breakable,enhanced,colback=white,arc=0pt,boxrule=0.5pt,boxsep=0pt,left=0pt,right=0pt,top=-3pt,bottom=-3pt,after={\vspace{-0.1cm}}]
        \begin{lstlisting}[label={code:patched-code},xleftmargin=-28pt,basicstyle=\scriptsize]
       6 public interface DSAEncoder {
       7     byte[] encode(BigInteger r, BigInteger s)
        \end{lstlisting}
        \end{tcolorbox}
        \caption{Source code snippet with repaired formatting.}
        \label{subfig:patched-code}
    \end{subfigure}
    \caption{\Styler from a Checkstyle formatting violation to a fix.}
    \label{fig:java-code}
\end{figure}

For that violation, \Styler encodes the incorrectly formatted lines (\autoref{subfig:violating-code}) into the abstract token sequence shown in \autoref{subfig:violating-code-tokenized}.
Then, this abstract token sequence is given as input to LSTM models, which predict alternative formatting token sequences, as the one shown in \autoref{subfig:patched-machine-learning-output}, that may fix the current formatting violation.
These predicted formatting token sequences are then used to modify the formatting tokens of the original abstract token sequence.
It results in predicted abstract token sequences, as the one shown in \autoref{subfig:patched-code-tokenized}.
The difference between \autoref{subfig:violating-code-tokenized} and \autoref{subfig:patched-code-tokenized} is the replacement of the formatting token \texttt{1\_NL} by \texttt{1\_SP}. 
This predicted repair means that the line break before the token ``\{'' should be replaced by a single space.
Then, the predicted abstract token sequence (\autoref{subfig:patched-code-tokenized}) is translated back to Java code (\autoref{subfig:patched-code}). Finally, when running Checkstyle on the new Java code, no Checkstyle violation is raised, meaning that \Styler successfully repaired the violation.

\subsection{Java source code encoding}\label{sec:java-source-code-encoding}

\Styler encodes Java source code into an abstract token sequence that is required to predict formatting changes.
An abstract token sequence is composed of pairs of abstract Java tokens and abstract formatting tokens.
\Styler represents each \textit{Java token} as an abstract token by keeping the value of the Java keywords, separators, and operators (e.g., \texttt{+} $\rightarrow$ \texttt{+}), and by replacing the other token kinds such as literals, comments, and identifiers by their types (e.g., \texttt{x} $\rightarrow$ \texttt{Identifier}). 
For each pair of subsequent Java tokens, \Styler creates an abstract \textit{formatting token}, which depends on the presence of a new line.
If there is no new line, \Styler counts the number of whitespace characters, and then represents it as \texttt{n\_SP} when the characters are spaces and \texttt{n\_TB} when the characters are tabulations, where \texttt{n} is the number of whitespaces characters (e.g., \texttt{\char32} $\rightarrow$ \texttt{1\_SP}). If there is no whitespace between two Java tokens (e.g., \texttt{x=}), \Styler adds \texttt{0\_None} between the two Java tokens.

If there are new lines between two Java tokens, \Styler first counts the number of new lines and represents it as \texttt{n\_NL}, where \texttt{n} is the number of new lines.
Then, \Styler calculates the indentation delta ($\Delta$), i.e., the indentation difference, between the line containing the first Java token and the line containing the second Java token.
Positive indentation deltas are represented by \texttt{$\Delta$\_ID} (indent), negative ones are represented by \texttt{$\Delta$\_DD} (dedent), and deltas equal to zero, i.e., no indentation change between the lines, are represented by the absence of an indentation delta representation.
The complete representation after the calculation of the number of new lines and the indentation delta is \texttt{n\_NL\_$\Delta$\_(ID|DD)\_(SP|TB)}. For instance, in \autoref{subfig:violating-code}, the new line between lines 7 and 8 is represented by \texttt{1\_NL\_4\_ID\_SP}, i.e., one new line and indentation delta +4.

\subsection{Self-supervised training data generation}\label{sec:training-data-generation}

\Styler does not use predefined templates for repairing formatting violations.
\Styler uses machine learning for inferring a model to repair formatting violations and, consequently, it needs training data.
One option would be to mine past commits from the project under consideration to collect training data. However, there might not exist enough data in the history of the project for training models.

Therefore, to have enough data for training, our key insight is to generate the training data in a self-supervised manner.
The idea is to modify violation-free Java files belonging to the project under analysis to trigger formatting rule violations. A similar idea has been explored by \cite{yasunaga2020graph}.
Then, one obtains a pair of files ($\alpha_{orig}$, $\alpha_{err}$): $\alpha_{orig}$ is the file without the formatting violation, and $\alpha_{err}$ is the file with the formatting violation. $\alpha_{orig}$ is a repaired version of $\alpha_{err}$, and we can use supervised machine learning to predict $\alpha_{orig}$ given $\alpha_{err}$.
We explore this idea in two different ways to generate training data, hereafter referred to as formatting violation injection \textit{protocols}. The protocol names are \protocolOne and \protocolTwo.

The \protocolOne protocol for injecting formatting violations in a project consists of automated insertion or deletion of a single formatting character (space, tabulation, or new line) in Java source files.
These modifications require a careful procedure so that 1) the project still compiles and 2) its behavior is not changed.
For this, we specify the locations in the source code files that are suitable to perform the modifications.
For insertions, the suitable locations are before or after any token.
For deletions, the suitable locations are 1) before or after any punctuation (``.'', ``,'', ``('', ``)'', ``['', ``]'', ``\{'', ``\}'', and ``;''), 2) before or after any operator (e.g., ``+'', ``-'', ``*'', ``='', ``+=''), and 3) in any token sequence longer than one indentation character.

The \protocolTwo protocol is meant to produce likely violations. Instead of directly changing the Java source code as \protocolOne, \protocolTwo performs modifications at the abstract token level.
The idea is to replace formatting tokens with the ones used by developers in the same context, i.e., between the same surrounding Java tokens.
For that, we use 3-grams, where a 3-gram = $\{$Java token, formatting token, Java token$\}$.
So given a violation-free Java file, the task of \protocolTwo is the following. First, the Java file is tokenized (see \autoref{sec:java-source-code-encoding}), and a random formatting token is picked and used to form a 3-gram, which is 3-gram$_{orig}$. Then, given a corpus of 3-grams previously created from software projects, \protocolTwo finds a 3-gram$_{i}$ in the corpus that matches the Java tokens of 3-gram$_{orig}$. Several matches can be found, but the selection of a 3-gram$_{i}$ is random according to its frequency in the corpus.
Then, the formatting token of 3-gram$_{orig}$ is replaced by the formatting token of 3-gram$_{i}$. Finally, \protocolTwo performs a de-tokenization so that a violating Java version of the original violation-free Java file is created.

\autoref{alg:injection_batch} presents the algorithm that \Styler uses to generate one training dataset per formatting violation injection protocol (\protocolOne and \protocolTwo).
The input of the algorithm is the format checker ruleset of the project, a corpus of violation-free Java files taken from the project, the number of violating files to be generated, the injection protocol to be used, and the maximum duration of the process.
Then, in each batch iteration (line 7), a file is randomly selected from the corpus of violation-free Java files (line 12), and the specified injection protocol is applied to it (line 13).
Once a batch is completed, the format checker is executed on the resulting modified files (line 16) so that the algorithm selects the ones that contain a single violation (line 17).
The algorithm ends when the desired number of files with violations is reached or when the process reaches the specified maximum duration.

\begin{algorithm}[t]
	\caption{Injection of formatting violations in Java files.}
    \label{alg:injection_batch}
    \begin{algorithmic}[1]
      \Require $ruleset$ -- format checker configuration of the project under consideration
      \Require $files$ -- corpus of violation-free Java files taken from the project
      \Require $numberOfViolations$ -- number of files with one violation to be generated 
      \Require $protocol$ in [\protocolOne, \protocolTwo]
      \Require $maxDuration$ -- maximum duration of the process
      \Ensure $dataset$ with Java files containing formatting violations raised by a format checker
      \State \textbf{var} $dataset \leftarrow \{\}$
      \State \textbf{var} $maxTime \leftarrow time.now + maxDuration$
      \State \textbf{var} $timeout \leftarrow false$
      \While{$dataset.size <$ $numberOfViolations$ \textbf{and not} $timeout$}
        \State \textbf{var} $modifiedFiles \leftarrow \{\}$
        \State batchSize $\leftarrow numberOfViolations - dataset.size$
        \For{$i \leftarrow 0$; $i < batchSize$; $i++$}
            \If{$time.now >= maxTime$}
                \State $timeout \leftarrow true$
                \State go to line 16
            \EndIf
            \State $file \leftarrow selectRandomly(files)$
            \State $file' \leftarrow changeFormatting(file, protocol)$
            \State $modifiedFiles.append(file')$
        \EndFor
        \State $formatCheckerResults \leftarrow runFormatChecker(modifiedFiles, ruleset)$
        \State $violatingFiles \leftarrow selectFilesWithOneViolation(formatCheckerResults)$
        \State $dataset.append(violatingFiles)$
      \EndWhile
      \State\Return $dataset$
    \end{algorithmic}
\end{algorithm}

\subsection{Violation encoding}\label{sec:violation-encoding}

In order to repair formatting violations, the Java source code encoded as an abstract token sequence must capture both the violation in the code and the context surrounding the violation.
So, for a given violation, \Styler considers a token window of $k$ source code lines before and after the violation location provided by the format checker for creating an abstract token sequence (see \autoref{sec:java-source-code-encoding}).
Once the violating line and the ones surrounding it are tokenized, \Styler places two tags around the tokens related to the origin of the violation so that the violation location and its type can be further identified.
The tags consist of the name of the format checker rule that was violated.
For instance, the violation presented in \autoref{fig:checkstyle-violation} is about the Checkstyle \emph{LeftCurly} rule, so the tags around the violation are \texttt{<LeftCurly>} and \texttt{</LeftCurly>} as shown in \autoref{subfig:violating-code-tokenized}.

The strategy to place the tags in the abstract token sequence is primarily based on the fact that the tags should surround the tokens related to the origin of the violation. At the same time, the number of tokens between the two tags should be minimal so to keep precise information about the violation location. Thus, \Styler places the tags according to the location information given by the format checker.
When the format checker provides the line and the column, \Styler places \texttt{<ViolationType>} one token before the violation and \texttt{</ViolationType>} one token after.
When the format checker provides the line but not the column (e.g., when the violation is about the Checkstyle \emph{LineLength} rule), \Styler places \texttt{<ViolationType>} one token before the line and \texttt{</ViolationType>} one token after the end of the line.

\subsection{Machine learning model}
\label{sec:model}

\noindent\textit{Learning (\autoref{fig:styler}--step C).}
\Styler aims to translate a token sequence with a formatting violation (input sequence) to a new token sequence with no formatting violation (output sequence).
\Styler uses a sequence-to-sequence translation based on recurrent neural network LSTMs (Long Short-Term Memory), similar to what is used for natural language translation.
Thanks to the token abstraction employed by \Styler to encode Java source code (see \autoref{sec:java-source-code-encoding} and \autoref{sec:violation-encoding}), the input and output vocabularies are small (respectively $\sim$150 and $\sim$50), hence are well handled by LSTM models.
\Styler uses LSTMs with bidirectional encoding, which means that the embedding is able to catch information around the formatting violation in the two directions. For instance, a violation triggered by the Checkstyle \emph{WhitespaceAround} rule, which checks that a token is surrounded by whitespaces, requires the contexts before and after the token.

\vspace{5pt}
\noindent\textit{Repairing (\autoref{fig:styler}--step F).} 
Once the LSTM models are trained (one per formatting violation injection protocol, see \autoref{sec:training-data-generation}), \Styler can be used for predicting fixes for a token sequence $I$ as in \autoref{subfig:violating-code-tokenized}. 
For an input sequence $I$, an LSTM model predicts $x$ alternative formatting token sequences using a technique called beam search, which we use off-the-shelf.
These alternatives are all potential repairs for the formatting violation (e.g., \autoref{subfig:patched-machine-learning-output}).

\begin{figure}[t]
    \centering
    \begin{subfigure}[b]{1\linewidth}
        \includegraphics[width=\columnwidth, frame]{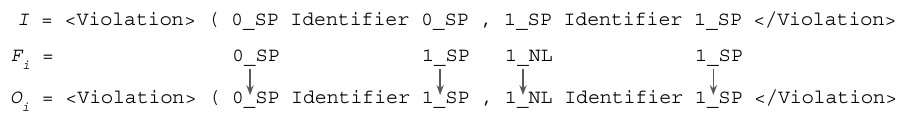}
        \caption{$length(F_i) = length(I)/2$.}
        \label{fig:mapping1}
    \end{subfigure}
    \begin{subfigure}[b]{1\linewidth}
        \vspace{2pt}
        \includegraphics[width=\columnwidth, frame]{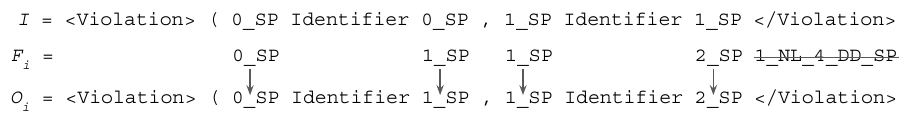}
        \caption{$length(F_i) > length(I)/2$.}
        \label{fig:mapping2}
    \end{subfigure}
    \begin{subfigure}[b]{1\linewidth}
        \vspace{2pt}
        \includegraphics[width=\columnwidth, frame]{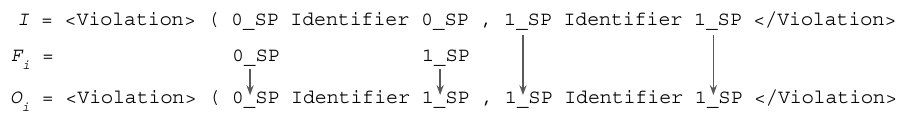}
        \caption{$length(F_i) < length(I)/2$.}
        \label{fig:mapping3}
    \end{subfigure}
    \caption{Generation of the sequence $O_i$ based on the predicted formatting tokens $F_i$ and the input $I$.}
    \label{fig:mappings}
\end{figure}

Note that the LSTM models predict \textit{formatting} token sequences (e.g., \autoref{subfig:patched-machine-learning-output}), but the goal is to have abstract token sequences containing Java and formatting tokens (e.g., \autoref{subfig:patched-code-tokenized}), so they can further be translated back to Java code.
For that, \Styler generates a new abstract token sequence ($O_i$) for each formatting token sequence ($F_i$), based on the original input $I$, such as in \autoref{fig:mapping1}.
Recall that $I$ is composed of pairs of Java tokens and formatting tokens (see \autoref{sec:java-source-code-encoding}), therefore its number of formatting tokens is $L_I = length(I)/2$.
However, an LSTM model does not enforce the output size, thus we cannot guarantee that the length of a predicted formatting token sequence ($L_{F_i} = length(F_i)$) is equal to $L_I$.
If $L_{F_i}>L_I$, \Styler uses the first $L_I$ formatting tokens from $F_i$ and ignores the remaining ones to generate $O_i$, such as in \autoref{fig:mapping2}.
If $L_{F_i}<L_I$, \Styler uses all formatting tokens from $F_i$ and copies the $L_{F_i}+1, L_{F_i}+2, \ldots, L_I$ original formatting tokens from $I$, such as in \autoref{fig:mapping3}.
Finally, after creating $x$ abstract token sequences $O$, \Styler continues its workflow (\autoref{fig:styler}--step G).

\subsection{Repair verification and selection}

\Styler performs $x$ predictions per LSTM model (i.e., \protocolOne-based model and \protocolTwo-based model), so in the end \Styler generates $x \times 2$ predictions to repair a single violation.
After the translation of these predictions back to Java source code (\autoref{fig:styler}--step G), \Styler performs a verification (\autoref{fig:styler}--step H), where the format checker is executed on the resulting Java source code files.
Finally, given the files that do not result in formatting violations, \Styler selects the one that has the smallest source code diff to give as output (\autoref{fig:styler}--step I).

\subsection{Implementation}\label{sec:implementation}

The approach employed by \Styler is independent of the considered format checker. The current implementation uses Checkstyle, which is a popular format checker for Java. Other format checkers can be integrated in \Styler. However, they must output the violation type and the violation location. This is necessary for the violation encoding (see \autoref{sec:violation-encoding}).

\Styler is implemented in Python. We use javalang\footnote{\url{https://github.com/c2nes/javalang/}, last access: 2020-07-13} for parsing and OpenNMT-py\footnote{\url{https://github.com/OpenNMT/OpenNMT-py/}, last access: 2020-07-13} for the machine learning part. \Styler is publicly available at \url{https://github.com/KTH/styler/}. The current calibration of \Styler is presented in \autoref{sec:styler-calibration}.

\section{Evaluation design}\label{sec:evaluation-design}

We conducted an empirical study to evaluate \Styler from different perspectives (see \autoref{sec:research-questions}), including a comparison against three state-of-the-art code formatting systems (see \autoref{sec:systems-to-compare}). We first built a dataset of Checkstyle violations mined from GitHub repositories (see \autoref{sec:data-collection}), and then we gave these violations as input to all the four tools (see \autoref{sec:setup}) to measure their repairability. In this section, we present the design of our study.

\subsection{Research questions}\label{sec:research-questions}

Our goal is to answer the following six research questions.

\vspace{7pt}
\newcommand\rqone{To what extent does \Styler repair real-world Checkstyle formatting violations, compared to other systems?}
\noindent \textbf{RQ \#1} [Overall repairability]: \rqone

\noindent Overall repairability is an important metric to measure the value of tools. We investigate the repairability of \Styler on real Checkstyle violations, which allows us to understand to what extent \Styler repairs formatting violations that have occurred in practice.
Moreover, we compare the repairability of \Styler to the repairability of three code formatters by using the same dataset of violations to investigate if, and to what extent, \Styler outperforms the related systems.

\vspace{7pt}
\newcommand\rqtwo{To what extent does \Styler repair different violation types, compared to other systems?}
\noindent \textbf{RQ \#2} [Violation-type-based repairability]: \rqtwo

\noindent Checkstyle has different formatting rules, so it raises different violation types. In this research question, we investigate if, and to what extent, \Styler repairs different violation types compared to the other systems. This analysis is also important to investigate if the systems are complementary to each other.

\vspace{7pt}
\newcommand\rqthree{What are the cases in which \Styler fails to generate a correct repair?}
\noindent \textbf{RQ \#3} [Unsuccessful repair cases]: \rqthree

\noindent Understanding the cases in which \Styler fails to generate a correct repair is important so that i) \Styler can be further improved, ii) hard-to-repair violations are identified and, consequently, researchers might study them and develop tools specialized to repair them, and iii) the limitations of \Styler can be taken into account by developers when deciding whether or not to use \Styler in their projects. To discover the most frequent cases in which \Styler does not succeed to generate a correct repair, we manually analyzed violations of the rules for which \Styler does not perform well.

\vspace{7pt}
\newcommand\rqfour{What is the size of the repairs generated by \Styler, compared to other systems?}
\noindent \textbf{RQ \#4} [Quality]: \rqfour

\noindent There may be several alternative repairs that fix a given Checkstyle violation, including ones that change source code lines other than the ill-formatted line. In this research question, we compare the size of the repairs generated by \Styler against the repairs generated by the other systems.

\vspace{7pt}
\newcommand\rqfive{How fast is \Styler for learning and predicting formatting repairs?}
\noindent \textbf{RQ \#5} [Performance]: \rqfive

\noindent To investigate if \Styler could be applicable in practice, we measure its performance for fixing the mined Checkstyle violations. This is valuable information for those who could be interested in using \Styler as a pre-commit hook in IDEs or continuous integration services.

\vspace{7pt}
\newcommand\rqsix{How do the two training data generation techniques of \Styler contribute to its repairability?}
\noindent \textbf{RQ \#6} [Technical analysis]: \rqsix

\noindent Finally, we perform a comparison between the two formatting violation injection protocols used to generate training data (see \autoref{sec:training-data-generation}). This comparison is done through the LSTM models trained with the two different training sets. We investigate if, and to what extent, one of the models contributes more to the repairability of \Styler. This is an important investigation from the point of view of users who might want to use \Styler with only one model for performance reasons.

\subsection{Systems under comparison}\label{sec:systems-to-compare}

We selected three systems to compare \Styler with: \CheckStyleIDEA \citep{checkstyle-idea}, \Naturalize \citep{Allamanis2014naturalize}, and \CodeBuff \citep{Parr2016codebuff}.
\CheckStyleIDEA, also referred to as \CSIDEA in this paper, is an IDE-based code formatter plugin for the IntelliJ IDE. It provides IDE-integrated feedback against a given Checkstyle ruleset and fixes Checkstyle violations through the IntelliJ formatter by taking into consideration a Checkstyle ruleset.
\Naturalize is a tool dedicated to assisting developers in fixing coding conventions related to naming and formatting in Java programs. It learns coding conventions from a codebase and suggests fixes to developers, such as formatting modifications, based on the n-gram model.
\CodeBuff is a code formatter applicable to any programming language with an ANTLR grammar. Instead of formatting the code according to ad-hoc rules for a language, \CodeBuff aims to infer the formatting rules given a grammar for the language and a set of files following the same formatting rules. For each token, a KNN model decides to indent it or to align it with another token based on the abstract syntax tree of the source file.

All three systems are code formatters. \CheckStyleIDEA takes a Checkstyle ruleset into consideration, and \Naturalize and \CodeBuff are the state-of-the-art machine learning formatters that aim to assist developers to fix formatting-related issues without any prior or ad-hoc formatting rules.

\subsection{Data collection}\label{sec:data-collection}

To execute \Styler and the systems under comparison and, consequently, answer our research questions, we created a dataset of Checkstyle formatting violations by mining open-source projects.
The first step was to build a list of projects, which was done based on the data previously collected for the study presented in \autoref{sec:study-checkstyle-usage}. We selected all the projects that have exactly one Checkstyle ruleset file and use Maven. This resulted in \numprint{2143} projects.

For each project, we tried to reproduce Checkstyle violations with the following automated lightweight approach.
First, the remote repository of the project is cloned from GitHub\footnote{All repositories were cloned on June 24--25, 2021.}. Then, a sanity check is performed on the \checkstylexml file contained in the project. If the file contains variables, the project is discarded. Otherwise, a search in the history of the project is done for the last commit ($c_i$) that contains modifications in the \checkstylexml file, which is the commit to be used as the starting point for the reproduction of real violations.
Then, $c_i$ is checked out, and all the files of the project are submitted to a process that aims to check if our automated approach can successfully execute Checkstyle on the project and with which version of Checkstyle. The latter is necessary because new versions of Checkstyle might introduce breaking backward compatibility\footnote{Checkstyle release notes: \url{https://checkstyle.sourceforge.io/releasenotes.html}} and, then, fail to parse a \checkstylexml file that was used with previous versions of Checkstyle. Such a process consists of executing multiple Checkstyle versions on the project, from a newer version to an older one, until finding one version that does not fail or until the available options end\footnote{Our current implementation supports 48 Checkstyle versions, from 8.0 to 8.43.}.
If a successful Checkstyle execution is found, the last tested Checkstyle version, $x$, is chosen to be used on the project. All commits since $c_i$ are then gathered, inclusive, so that all commits to be analyzed are based on the same Checkstyle ruleset.

Then, each selected commit is checked out, and a sanity check is performed on the \texttt{pom.xml} file of the commit. If it points to a suppression file, the commit is discarded because we want violations that happened in practice and our lightweight approach does not solve paths. Otherwise, the Checkstyle version $x$ is executed on the files of the project. If the commit under analysis is the first one to be analyzed, Checkstyle is executed on all the files of the project. Otherwise, Checkstyle is executed only on the files changed in the commit under analysis to avoid duplicate violations in the dataset. Moreover, Java files in folders named \textit{test} or \textit{resources} are ignored since we want violations that happened in the main source code. Then, after executing Checkstyle, if at least one Checkstyle violation is raised, the violating Java files and information about the violations, e.g., Checkstyle violation types and locations, are saved.

Such a process was executed for all the \numprint{2143} projects in our list.
At the end of the process, we removed duplicate Java files according to the file content among all commits if any.
Then, we selected the files containing a single Checkstyle violation that is related to formatting. We performed this selection to accurately evaluate repairs produced by \Styler and the other tools. Finally, we kept projects where all criteria yield at least 20 Checkstyle formatting violations.
By applying this systematic reproduction and selection process, we obtained a dataset containing \nbViolationsTotalToPrint Checkstyle violations spread over \nbProjectsTotalToPrint projects. We used one project, and the violations found in it (\nbViolationsCalibrationToPrint), to calibrate \Styler (see \autoref{sec:styler-calibration}), and the other \nbProjectsEvaluationToPrint projects with \nbViolationsEvaluationToPrint violations for the actual evaluation.

The dataset is diverse in terms of projects and violation types.
\autoref{tab:real-violation-dataset-projects} shows the subject projects. The projects are very diverse in several aspects, such as in number of Java files ($min = 2$, $med = 532$, $max = 18,206$), commits ($min = 13$, $med = 1,251$, $max = 71,544$), contributors ($min = 1$, $med = 14$, $max = 249$), and stars ($min = 5$, $med = 46$, $max = 24,888$).
Additionally, \autoref{tab:real-violation-dataset} shows the stats per Checkstyle formatting rule. We note that the most frequent violations in our dataset are violations of the rules \emph{RegexpSingleline}, \emph{EmptyLineSeparator}, and \emph{LineLength}.

\begin{small}
\begin{longtable}{@{}p{.457\textwidth} r r r r r@{}}
    \caption[Subject projects. \nbProjectsEvaluationToPrint projects were used to evaluate \Styler and one project (flagged with ``$\star$'') was used to calibrate \Styler.]{Subject projects\footnote{The numbers of commits, contributors, and stars were retrieved on June 23--24, 2021. The number of violations is the number of formatting-related violations found in the project and that were selected according to our criteria.}. \nbProjectsEvaluationToPrint projects were used to evaluate \Styler and one project (flagged with ``$\star$'') was used to calibrate \Styler.}
    \label{tab:real-violation-dataset-projects} \\

    \toprule
    Projects (\nbProjectsEvaluationToPrint + 1) & \rotatebox{70}{\parbox{1cm}{\centering \# Java files}} & \rotatebox{70}{\parbox{1.2cm}{\centering \# Commits}} & \rotatebox{70}{\parbox{1.2cm}{\centering \scriptsize \# Contributors}} & \rotatebox{70}{\parbox{1.2cm}{\centering \# Stars}} & \rotatebox{70}{\parbox{1.2cm}{\centering \# Violations}} \\
    \midrule
    \endfirsthead
    
    \toprule
    \multicolumn{6}{@{}l}{\textit{continued from previous page}}\\
    \midrule
    Projects (\nbProjectsEvaluationToPrint + 1) & \rotatebox{70}{\parbox{1cm}{\centering \# Java files}} & \rotatebox{70}{\parbox{1.2cm}{\centering \# Commits}} & \rotatebox{70}{\parbox{1.2cm}{\centering \scriptsize \# Contributors}} & \rotatebox{70}{\parbox{1.2cm}{\centering \# Stars}} & \rotatebox{70}{\parbox{1.2cm}{\centering \# Violations}} \\
    \midrule
    \endhead
    
    \midrule
    \multicolumn{6}{r@{}}{\textit{continued on next page}}\\
    \bottomrule
    \endfoot
    
    \bottomrule
    \endlastfoot
    
1and1/cosmo & \numprint{557} & \numprint{594} & \numprint{8} & \numprint{54} & \numprint{26} \\
actiontech/txle & \numprint{208} & \numprint{1416} & \numprint{44} & \numprint{35} & \numprint{49} \\
Activiti/Activiti & \numprint{1532} & \numprint{10758} & \numprint{192} & \numprint{7837} & \numprint{1032} \\
Angel-ML/angel & \numprint{861} & \numprint{2714} & \numprint{47} & \numprint{6282} & \numprint{1143} \\
apache/crunch & \numprint{522} & \numprint{1093} & \numprint{34} & \numprint{100} & \numprint{26} \\
apache/ignite-3 & \numprint{344} & \numprint{201} & \numprint{23} & \numprint{23} & \numprint{80} \\
apache/iotdb & \numprint{1215} & \numprint{5001} & \numprint{121} & \numprint{1388} & \numprint{491} \\
apache/metron & \numprint{568} & \numprint{1453} & \numprint{60} & \numprint{828} & \numprint{115} \\
apache/servicecomb-java-chassis & \numprint{907} & \numprint{2997} & \numprint{98} & \numprint{1703} & \numprint{398} \\
apache/shardingsphere & \numprint{1441} & \numprint{29153} & \numprint{249} & \numprint{14097} & \numprint{43} \\
apache/usergrid & \numprint{639} & \numprint{10954} & \numprint{78} & \numprint{992} & \numprint{145} \\
\scriptsize{aspose-words-cloud/aspose-words-cloud-java} & \numprint{313} & \numprint{325} & \numprint{8} & \numprint{14} & \numprint{594} \\
atlanmod/NeoEMF & \numprint{293} & \numprint{2998} & \numprint{10} & \numprint{38} & \numprint{87} \\
bakdata/conquery & \numprint{628} & \numprint{7831} & \numprint{17} & \numprint{25} & \numprint{884} \\
benetech/ServiceNet & \numprint{660} & \numprint{1618} & \numprint{12} & \numprint{7} & \numprint{69} \\
black-ant/case & \numprint{1899} & \numprint{353} & \numprint{2} & \numprint{31} & \numprint{46} \\
blockchain-lab/ScaleOutDistributedLedger & \numprint{2} & \numprint{574} & \numprint{6} & \numprint{10} & \numprint{29} \\
ByteHamster/PSE & \numprint{133} & \numprint{835} & \numprint{5} & \numprint{8} & \numprint{29} \\
CESNET/perun & \numprint{1500} & \numprint{7359} & \numprint{42} & \numprint{46} & \numprint{525} \\
Chaklader/Multi-threading & \numprint{532} & \numprint{55} & \numprint{1} & \numprint{16} & \numprint{35} \\
Chaklader/Object-Oriented-Design & \numprint{2985} & \numprint{35} & \numprint{1} & \numprint{240} & \numprint{67} \\
chenpengliang0909/WxJava & \numprint{622} & \numprint{2209} & \numprint{73} & \numprint{13} & \numprint{30} \\
ciandt-dev/tech-gallery & \numprint{122} & \numprint{1312} & \numprint{20} & \numprint{38} & \numprint{20} \\
cloudera/director-sdk & \numprint{516} & \numprint{39} & \numprint{7} & \numprint{20} & \numprint{317} \\
codefollower/H2-Research & \numprint{656} & \numprint{12913} & \numprint{114} & \numprint{387} & \numprint{573} \\
CONNECT-Solution/CONNECT & \numprint{1860} & \numprint{11828} & \numprint{33} & \numprint{65} & \numprint{1463} \\
couchbase/couchbase-java-client & \numprint{164} & \numprint{1051} & \numprint{22} & \numprint{257} & \numprint{28} \\
couchbase/couchbase-jvm-core & \numprint{242} & \numprint{968} & \numprint{17} & \numprint{31} & \numprint{98} \\
ctripcorp/apollo & \numprint{78} & \numprint{2526} & \numprint{78} & \numprint{24888} & \numprint{144} \\
DaGeRe/peass & \numprint{103} & \numprint{907} & \numprint{1} & \numprint{7} & \numprint{114} \\
decorators-squad/eo-yaml & \numprint{58} & \numprint{834} & \numprint{12} & \numprint{203} & \numprint{39} \\
decorators-squad/versioneye-api & \numprint{4} & \numprint{127} & \numprint{3} & \numprint{7} & \numprint{57} \\
delight-im/NationSoccer & \numprint{847} & \numprint{18} & \numprint{1} & \numprint{13} & \numprint{20} \\
DigitalMediaServer/DigitalMediaServer & \numprint{502} & \numprint{7315} & \numprint{38} & \numprint{24} & \numprint{42} \\
DSC-SPIDAL/harp & \numprint{580} & \numprint{900} & \numprint{18} & \numprint{20} & \numprint{50} \\
dzhw/metadatamanagement & \numprint{528} & \numprint{8979} & \numprint{18} & \numprint{21} & \numprint{163} \\
eclipse-ee4j/glassfish & \numprint{12820} & \numprint{1337} & \numprint{48} & \numprint{233} & \numprint{140} \\
eclipse/milo & \numprint{1434} & \numprint{1119} & \numprint{16} & \numprint{618} & \numprint{273} \\
EMResearch/EMB & \numprint{1461} & \numprint{268} & \numprint{3} & \numprint{5} & \numprint{112} \\
fangjinuo/easyjson & \numprint{91} & \numprint{624} & \numprint{3} & \numprint{58} & \numprint{68} \\
farao-community/farao-core & \numprint{230} & \numprint{364} & \numprint{17} & \numprint{7} & \numprint{51} \\
findbugsproject/findbugs & \numprint{2310} & \numprint{15375} & \numprint{35} & \numprint{661} & \numprint{35} \\
GluuFederation/casa & \numprint{140} & \numprint{698} & \numprint{9} & \numprint{9} & \numprint{61} \\
GluuFederation/oxCore & \numprint{353} & \numprint{1457} & \numprint{17} & \numprint{11} & \numprint{201} \\
gomint/gomint & \numprint{2953} & \numprint{1668} & \numprint{26} & \numprint{221} & \numprint{42} \\
googleapis/google-cloud-java & \numprint{65} & \numprint{5128} & \numprint{200} & \numprint{1559} & \numprint{843} \\
GourdErwa/MyNote & \numprint{1151} & \numprint{53} & \numprint{1} & \numprint{72} & \numprint{57} \\
graphfoundation/ongdb & \numprint{4587} & \numprint{220} & \numprint{2} & \numprint{211} & \numprint{4569} \\
griddynamics/jagger & \numprint{870} & \numprint{2343} & \numprint{25} & \numprint{65} & \numprint{184} \\
Gurux/gurux.dlms.java & \numprint{175} & \numprint{377} & \numprint{2} & \numprint{52} & \numprint{31} \\
h2database/h2database & \numprint{656} & \numprint{13570} & \numprint{127} & \numprint{2849} & \numprint{215} \\
HealerJean/HealerJean.github.io & \numprint{2752} & \numprint{1385} & \numprint{2} & \numprint{23} & \numprint{112} \\
HuygensING/timbuctoo & \numprint{621} & \numprint{8311} & \numprint{14} & \numprint{38} & \numprint{66} \\
ibinti/bugvm & \numprint{9665} & \numprint{3647} & \numprint{23} & \numprint{348} & \numprint{171} \\
$\star$ inovexcorp/mobi & \numprint{599} & \numprint{12454} & \numprint{19} & \numprint{35} & \numprint{267} \\
Internet2/grouper & \numprint{4661} & \numprint{10368} & \numprint{22} & \numprint{59} & \numprint{101} \\
intuit/Tank & \numprint{1160} & \numprint{320} & \numprint{7} & \numprint{61} & \numprint{217} \\
IQSS/dataverse & \numprint{807} & \numprint{19817} & \numprint{119} & \numprint{600} & \numprint{132} \\
ita-social-projects/GreenCity & \numprint{182} & \numprint{2164} & \numprint{38} & \numprint{41} & \numprint{29} \\
java110/MicroCommunity & \numprint{711} & \numprint{4172} & \numprint{12} & \numprint{516} & \numprint{584} \\
jflex-de/jflex & \numprint{126} & \numprint{1874} & \numprint{11} & \numprint{400} & \numprint{53} \\
junkdog/artemis-odb & \numprint{266} & \numprint{1754} & \numprint{26} & \numprint{638} & \numprint{111} \\
kitodo/kitodo-production & \numprint{602} & \numprint{12109} & \numprint{30} & \numprint{47} & \numprint{43} \\
ldp4j/ldp4j & \numprint{752} & \numprint{459} & \numprint{2} & \numprint{43} & \numprint{40} \\
liuawen/Learning-Java & \numprint{18206} & \numprint{312} & \numprint{2} & \numprint{184} & \numprint{20} \\
LoboEvolution/LoboEvolution & \numprint{1712} & \numprint{416} & \numprint{1} & \numprint{25} & \numprint{580} \\
ManfredTremmel/gwt-bean-validators & \numprint{934} & \numprint{219} & \numprint{3} & \numprint{26} & \numprint{23} \\
matsim-org/matsim-episim-libs & \numprint{59} & \numprint{2528} & \numprint{18} & \numprint{15} & \numprint{33} \\
NationalSecurityAgency/datawave & \numprint{1939} & \numprint{2343} & \numprint{42} & \numprint{360} & \numprint{140} \\
NationalSecurityAgency/emissary & \numprint{317} & \numprint{186} & \numprint{16} & \numprint{171} & \numprint{20} \\
neo4j/neo4j & \numprint{3641} & \numprint{71544} & \numprint{210} & \numprint{9060} & \numprint{3375} \\
neuhalje/bouncy-gpg & \numprint{44} & \numprint{440} & \numprint{8} & \numprint{168} & \numprint{44} \\
O2-Czech-Republic/proxima-platform & \numprint{314} & \numprint{1611} & \numprint{7} & \numprint{12} & \numprint{91} \\
omnisci/omnisci-jdbc & \numprint{92} & \numprint{76} & \numprint{12} & \numprint{6} & \numprint{61} \\
open-eid/digidoc4j & \numprint{287} & \numprint{2000} & \numprint{25} & \numprint{55} & \numprint{95} \\
opencb/opencga & \numprint{837} & \numprint{12047} & \numprint{39} & \numprint{120} & \numprint{410} \\
OpenEMS/openems & \numprint{1615} & \numprint{4962} & \numprint{32} & \numprint{194} & \numprint{46} \\
openmessaging/openmessaging-java & \numprint{55} & \numprint{332} & \numprint{11} & \numprint{696} & \numprint{61} \\
Plugily-Projects/MurderMystery & \numprint{66} & \numprint{694} & \numprint{15} & \numprint{105} & \numprint{105} \\
ppati000/visualDFA & \numprint{18} & \numprint{270} & \numprint{6} & \numprint{8} & \numprint{26} \\
primefaces/primefaces & \numprint{1250} & \numprint{12177} & \numprint{160} & \numprint{1367} & \numprint{33} \\
Qihoo360/Quicksql & \numprint{1329} & \numprint{191} & \numprint{11} & \numprint{1733} & \numprint{109} \\
Roboy/roboy\_dialog & \numprint{419} & \numprint{1251} & \numprint{17} & \numprint{9} & \numprint{26} \\
Rugal/algorithm & \numprint{97} & \numprint{421} & \numprint{1} & \numprint{6} & \numprint{51} \\
RWTH-i5-IDSG/steve & \numprint{185} & \numprint{1375} & \numprint{11} & \numprint{215} & \numprint{174} \\
seedstack/business & \numprint{252} & \numprint{309} & \numprint{7} & \numprint{20} & \numprint{196} \\
seedstack/seed & \numprint{439} & \numprint{698} & \numprint{10} & \numprint{24} & \numprint{351} \\
self-xdsd/self-core & \numprint{222} & \numprint{1488} & \numprint{11} & \numprint{19} & \numprint{52} \\
self-xdsd/self-storage & \numprint{21} & \numprint{384} & \numprint{5} & \numprint{11} & \numprint{35} \\
SentinelDataHub/dhus-core & \numprint{412} & \numprint{13} & \numprint{3} & \numprint{6} & \numprint{35} \\
SergeyZhernovoy/Java-education & \numprint{265} & \numprint{441} & \numprint{1} & \numprint{8} & \numprint{40} \\
spark-root/laurelin & \numprint{44} & \numprint{440} & \numprint{4} & \numprint{7} & \numprint{30} \\
StevenLooman/sonar-magik & \numprint{54} & \numprint{178} & \numprint{2} & \numprint{5} & \numprint{27} \\
Stratio/bdt & \numprint{54} & \numprint{1023} & \numprint{34} & \numprint{65} & \numprint{70} \\
StuPro-TOSCAna/TOSCAna & \numprint{122} & \numprint{606} & \numprint{6} & \numprint{10} & \numprint{33} \\
TIBCOSoftware/genxdm & \numprint{1025} & \numprint{930} & \numprint{5} & \numprint{6} & \numprint{501} \\
tmobile/kardio & \numprint{206} & \numprint{41} & \numprint{9} & \numprint{149} & \numprint{130} \\
toast-tk/toast-tk-engine & \numprint{207} & \numprint{418} & \numprint{5} & \numprint{12} & \numprint{41} \\
twilio/twilio-java & \numprint{654} & \numprint{2028} & \numprint{93} & \numprint{377} & \numprint{1126} \\
V1toss/JavaPA & \numprint{146} & \numprint{242} & \numprint{1} & \numprint{16} & \numprint{28} \\
vassalengine/vassal & \numprint{986} & \numprint{7861} & \numprint{14} & \numprint{99} & \numprint{28} \\
vostok/hercules & \numprint{322} & \numprint{1988} & \numprint{9} & \numprint{16} & \numprint{118} \\
wayshall/onetwo & \numprint{1221} & \numprint{4690} & \numprint{2} & \numprint{17} & \numprint{158} \\
wso2-attic/commons & \numprint{7399} & \numprint{904} & \numprint{1} & \numprint{10} & \numprint{587} \\
zanata/zanata-platform & \numprint{1743} & \numprint{19064} & \numprint{29} & \numprint{288} & \numprint{70} \\
\midrule
min & \numprint{2} & \numprint{13} & \numprint{1} & \numprint{5} & \numprint{20} \\
med & \numprint{532} & \numprint{1251} & \numprint{14} & \numprint{46} & \numprint{70} \\
max & \numprint{18206} & \numprint{71544} & \numprint{249} & \numprint{24888} & \numprint{4569} \\

\end{longtable}
\end{small}

\vspace{-16pt}
\begin{table}[ht]
    \caption[Checkstyle formatting violation dataset stats per rule.]{Checkstyle formatting violation dataset stats per rule\footnote{Considering Checkstyle, \Styler also targets the following rules that are not contained in our dataset: \emph{AnnotationLocation}, \emph{AnnotationOnSameLine}, \emph{EmptyForInitializerPad}, and \emph{TypecastParenPad}.}.}
    \label{tab:real-violation-dataset}
    \vspace{-6pt}
    \centering
    \begin{tabular}{@{}l R{.06\textwidth} @{\hskip 0.05in} L{.05\textwidth} @{\hskip 0.1in} L{.06\textwidth} @{\hskip 0.15in} R{.1\textwidth} @{\hskip 0.05in} L{.05\textwidth} @{\hskip 0.1in} L{.06\textwidth} @{\hskip 0.1in} R{.05\textwidth} @{\hskip 0.05in} L{.05\textwidth} @{\hskip 0.1in} L{.06\textwidth} @{\hskip 0.1in}@{}}
        \toprule
        \multirow{2}{*}{Checkstyle rule (25)} & \multicolumn{3}{c}{Calibration} & \multicolumn{6}{c}{Evaluation} \\
        {} & \multicolumn{3}{c}{Violations (\nbViolationsCalibrationToPrint)} & \multicolumn{3}{c}{Violations (\nbViolationsEvaluationToPrint)} & \multicolumn{3}{c}{Projects (\nbProjectsEvaluationToPrint)} \\
        \midrule

CommentsIndentation & \ChartWithAbsoluteAndPercentageNumbers{12}{\nbViolationsCalibration} & \ChartWithAbsoluteAndPercentageNumbers{30}{\nbViolationsEvaluation} & \ChartWithAbsoluteAndPercentageNumbers{10}{\nbProjectsEvaluation} \\
EmptyForIteratorPad & \ChartWithAbsoluteAndPercentageNumbers{0}{\nbViolationsCalibration} & \ChartWithAbsoluteAndPercentageNumbers{11}{\nbViolationsEvaluation} & \ChartWithAbsoluteAndPercentageNumbers{2}{\nbProjectsEvaluation} \\
EmptyLineSeparator & \ChartWithAbsoluteAndPercentageNumbers{11}{\nbViolationsCalibration} & \ChartWithAbsoluteAndPercentageNumbers{4837}{\nbViolationsEvaluation} & \ChartWithAbsoluteAndPercentageNumbers{34}{\nbProjectsEvaluation} \\
FileTabCharacter & \ChartWithAbsoluteAndPercentageNumbers{4}{\nbViolationsCalibration} & \ChartWithAbsoluteAndPercentageNumbers{1807}{\nbViolationsEvaluation} & \ChartWithAbsoluteAndPercentageNumbers{30}{\nbProjectsEvaluation} \\
GenericWhitespace & \ChartWithAbsoluteAndPercentageNumbers{3}{\nbViolationsCalibration} & \ChartWithAbsoluteAndPercentageNumbers{4}{\nbViolationsEvaluation} & \ChartWithAbsoluteAndPercentageNumbers{4}{\nbProjectsEvaluation} \\
Indentation & \ChartWithAbsoluteAndPercentageNumbers{7}{\nbViolationsCalibration} & \ChartWithAbsoluteAndPercentageNumbers{1531}{\nbViolationsEvaluation} & \ChartWithAbsoluteAndPercentageNumbers{23}{\nbProjectsEvaluation} \\
LeftCurly & \ChartWithAbsoluteAndPercentageNumbers{0}{\nbViolationsCalibration} & \ChartWithAbsoluteAndPercentageNumbers{400}{\nbViolationsEvaluation} & \ChartWithAbsoluteAndPercentageNumbers{31}{\nbProjectsEvaluation} \\
LineLength & \ChartWithAbsoluteAndPercentageNumbers{142}{\nbViolationsCalibration} & \ChartWithAbsoluteAndPercentageNumbers{4270}{\nbViolationsEvaluation} & \ChartWithAbsoluteAndPercentageNumbers{59}{\nbProjectsEvaluation} \\
MethodParamPad & \ChartWithAbsoluteAndPercentageNumbers{0}{\nbViolationsCalibration} & \ChartWithAbsoluteAndPercentageNumbers{86}{\nbViolationsEvaluation} & \ChartWithAbsoluteAndPercentageNumbers{13}{\nbProjectsEvaluation} \\
NoLineWrap & \ChartWithAbsoluteAndPercentageNumbers{0}{\nbViolationsCalibration} & \ChartWithAbsoluteAndPercentageNumbers{30}{\nbViolationsEvaluation} & \ChartWithAbsoluteAndPercentageNumbers{2}{\nbProjectsEvaluation} \\
NoWhitespaceAfter & \ChartWithAbsoluteAndPercentageNumbers{0}{\nbViolationsCalibration} & \ChartWithAbsoluteAndPercentageNumbers{59}{\nbViolationsEvaluation} & \ChartWithAbsoluteAndPercentageNumbers{9}{\nbProjectsEvaluation} \\
NoWhitespaceBefore & \ChartWithAbsoluteAndPercentageNumbers{0}{\nbViolationsCalibration} & \ChartWithAbsoluteAndPercentageNumbers{179}{\nbViolationsEvaluation} & \ChartWithAbsoluteAndPercentageNumbers{20}{\nbProjectsEvaluation} \\
OneStatementPerLine & \ChartWithAbsoluteAndPercentageNumbers{0}{\nbViolationsCalibration} & \ChartWithAbsoluteAndPercentageNumbers{4}{\nbViolationsEvaluation} & \ChartWithAbsoluteAndPercentageNumbers{2}{\nbProjectsEvaluation} \\
OperatorWrap & \ChartWithAbsoluteAndPercentageNumbers{10}{\nbViolationsCalibration} & \ChartWithAbsoluteAndPercentageNumbers{274}{\nbViolationsEvaluation} & \ChartWithAbsoluteAndPercentageNumbers{26}{\nbProjectsEvaluation} \\
ParenPad & \ChartWithAbsoluteAndPercentageNumbers{0}{\nbViolationsCalibration} & \ChartWithAbsoluteAndPercentageNumbers{110}{\nbViolationsEvaluation} & \ChartWithAbsoluteAndPercentageNumbers{12}{\nbProjectsEvaluation} \\
Regexp & \ChartWithAbsoluteAndPercentageNumbers{0}{\nbViolationsCalibration} & \ChartWithAbsoluteAndPercentageNumbers{688}{\nbViolationsEvaluation} & \ChartWithAbsoluteAndPercentageNumbers{8}{\nbProjectsEvaluation} \\
RegexpMultiline & \ChartWithAbsoluteAndPercentageNumbers{0}{\nbViolationsCalibration} & \ChartWithAbsoluteAndPercentageNumbers{15}{\nbViolationsEvaluation} & \ChartWithAbsoluteAndPercentageNumbers{1}{\nbProjectsEvaluation} \\
RegexpSingleline & \ChartWithAbsoluteAndPercentageNumbers{0}{\nbViolationsCalibration} & \ChartWithAbsoluteAndPercentageNumbers{8678}{\nbViolationsEvaluation} & \ChartWithAbsoluteAndPercentageNumbers{20}{\nbProjectsEvaluation} \\
RegexpSinglelineJava & \ChartWithAbsoluteAndPercentageNumbers{0}{\nbViolationsCalibration} & \ChartWithAbsoluteAndPercentageNumbers{588}{\nbViolationsEvaluation} & \ChartWithAbsoluteAndPercentageNumbers{4}{\nbProjectsEvaluation} \\
RightCurly & \ChartWithAbsoluteAndPercentageNumbers{57}{\nbViolationsCalibration} & \ChartWithAbsoluteAndPercentageNumbers{540}{\nbViolationsEvaluation} & \ChartWithAbsoluteAndPercentageNumbers{23}{\nbProjectsEvaluation} \\
SeparatorWrap & \ChartWithAbsoluteAndPercentageNumbers{3}{\nbViolationsCalibration} & \ChartWithAbsoluteAndPercentageNumbers{7}{\nbViolationsEvaluation} & \ChartWithAbsoluteAndPercentageNumbers{3}{\nbProjectsEvaluation} \\
SingleSpaceSeparator & \ChartWithAbsoluteAndPercentageNumbers{0}{\nbViolationsCalibration} & \ChartWithAbsoluteAndPercentageNumbers{22}{\nbViolationsEvaluation} & \ChartWithAbsoluteAndPercentageNumbers{2}{\nbProjectsEvaluation} \\
TrailingComment & \ChartWithAbsoluteAndPercentageNumbers{0}{\nbViolationsCalibration} & \ChartWithAbsoluteAndPercentageNumbers{552}{\nbViolationsEvaluation} & \ChartWithAbsoluteAndPercentageNumbers{5}{\nbProjectsEvaluation} \\
WhitespaceAfter & \ChartWithAbsoluteAndPercentageNumbers{0}{\nbViolationsCalibration} & \ChartWithAbsoluteAndPercentageNumbers{1053}{\nbViolationsEvaluation} & \ChartWithAbsoluteAndPercentageNumbers{29}{\nbProjectsEvaluation} \\
WhitespaceAround & \ChartWithAbsoluteAndPercentageNumbers{18}{\nbViolationsCalibration} & \ChartWithAbsoluteAndPercentageNumbers{1016}{\nbViolationsEvaluation} & \ChartWithAbsoluteAndPercentageNumbers{50}{\nbProjectsEvaluation} \\

        \bottomrule
    \end{tabular}
\end{table}

\subsection{Setup and execution of the systems}\label{sec:setup}

We gave the dataset of violations as input to \Styler and the three systems under comparison to evaluate their repairability. In this section, we present the setup of the systems, which includes the calibration of \Styler, the adaptations performed in \Naturalize and \CodeBuff, and how the four systems were executed.

\subsubsection{\Styler calibration}\label{sec:styler-calibration}

To calibrate \Styler, i.e., the \protocolOne- and \protocolTwo-based models, we performed an exploratory study by training LSTM models with different configurations. The configurations combine values for key parameters, which are the model attention type (general or mlp), the number of layers (one or two) and units (256 or 512) for the model encoder/decoder, and the model word embedding size (256 or 512). For each configuration, the training was performed for a maximum of 20k iterations, with a batch size of 32, and a model was saved in the iterations 5k, 10k, 15k, and 20k. This means that, in the end, we obtained 64 models (2 model attention types $\times$ 2 numbers of layers $\times$ 2 numbers of units $\times$ 2 embedding sizes $\times$ 4 training iterations) per training data generation protocol (i.e., \protocolOne and \protocolTwo).

Those models were created for one open-source project\footnote{\url{https://github.com/inovexcorp/mobi}} contained in our dataset (see \autoref{sec:data-collection}), which was randomly selected from the top-5 projects with the most diversity of violated formatting rules. The project was given as input to \Styler, which produced training data by injecting Checkstyle violations in violation-free files found in the project (see \autoref{sec:training-data-generation}). For each protocol, 10k violations were injected. This data was used to train the LSTM models, where 9k violations were used for training and 1k for validation. Once the 64 models per protocol were created, we executed \Styler with each of them on the real violations found in the project so that we could test the models and choose the configuration of the best ones. Then, for each protocol, we picked the configuration of the model that repaired violations in a more balanced way in terms of Checkstyle rules. The best \protocolOne-based model was with the mlp model attention, one layer, 256 units, embedding size of 512, and 5k training iterations, and the best \protocolTwo-based model was with the same values for the numbers of layers, embedding size, and training iterations, but with the general model attention and 512 units. These are the configurations we used for training the models for our experiments presented in \autoref{sec:evaluation-results}.

For prediction, the beam search creates $x = 5$ potential repairs per model. As for the violation encoding (see \autoref{sec:violation-encoding}), we set $k = 6$. Recall that this parameter is about the token window before and after the violation (i.e., the context surrounding the violation). This parameter is made big enough to contain important information and, at the same time, small enough to still allow learning and prediction, and was set based on meta-optimization.

\subsubsection{\Naturalize and \CodeBuff adaptation}
To use \Naturalize, we had to slightly modify it. \Naturalize recommends multiple fixes, so we take the first one for a given violation as being the repair. In addition, we changed \Naturalize to only work for indentation, excluding fixes regarding naming conventions (which are out of the scope of this paper).
To run \CodeBuff, we give it the required configuration, including the number of spaces for indentation. This value depends on the project given as input to \CodeBuff. Thus, before running \CodeBuff on a project, we count the most frequent indentation size found in the violation-free files of the project and provide it to \CodeBuff.

\subsubsection{Execution of the systems}\label{sec:execution}

The four systems were executed to repair the \nbViolationsEvaluationToPrint violations found in the \nbProjectsEvaluationToPrint projects contained in the real violation dataset.
The machine-learning-based systems (\Styler, \Naturalize, and \CodeBuff) require a corpus of violation-free files to be trained.
Therefore, for each subject project, we selected, as training seeds, all violation-free Java files from the first commit, or any subsequent one, that uses the same Checkstyle ruleset used to collect the real violations.
We took special care of consistency in our experiment: all the three machine-learning-based systems were trained to repair a given project using the same corpus of violation-free files from the project.

\Styler requires other input for training. Recall that its training process includes a step for creating the actual training data (see \autoref{fig:styler}--step A), which is based on the corpus of violation-free files. For each protocol, we set \autoref{alg:injection_batch} to create \numprint{10000} files per project, with a maximum duration of three hours. The resulting files with violations were split for learning and validation in a balanced way according to the violation types, considering 90\% for learning and 10\% for validation.

Finally, to run \CheckStyleIDEA on each subject project, we first loaded the violating Java files and the \checkstylexml file contained in the project in IntelliJ. Then, we imported the Checkstyle ruleset \textit{(Settings > Editor > Code Style > Import Scheme > Checkstyle Configuration)} and simply called the function ``Reformat Code'' from the IDE.

\section{Evaluation results}\label{sec:evaluation-results}

We present and discuss the results for our six research questions in this section.

\subsection{Overall repairability (RQ \#1)}

To investigate the overall repairability of \Styler and the other three systems on the \nbViolationsEvaluationToPrint Checkstyle violations, we categorized the repair attempts per status, as shown in \autoref{tab:rq1}. There are two groups of status: \textit{repaired} and \textit{not repaired}. The repaired violations are either \textit{fully repaired}, i.e., no violation is raised after the repair attempt, or partially repaired, i.e., the violation no longer exists in the source code but \textit{new violations} were introduced. For the sake of clarity, it is worth mentioning that only the full repairs are used for the other five research questions. The group of violations that were not repaired includes the cases where a resulting source code file still contains the \textit{same violation} only or the \textit{same + new} violations, or is \textit{broken}, which means that the file cannot be parsed by javalang after the repair attempt.

\Styler fully repaired 41\% of the violations while \CSIDEA repaired 50\%, which is the greatest overall repairability among the four considered tools.
\Naturalize and \CodeBuff repaired fewer violations (15\% and 20\%, respectively). To check if there is a significant difference between \Styler and the other tools regarding the full repairs, we used McNemar test. \autoref{tab:rq1-contingency-tables} shows the contingency tables given as input to the test. We found \textit{p-value} $<$ 0.00001 for all the three tests. Considering $\alpha=0.05$, this means that \Styler and any other tool have a statistically significant different proportion of errors on our dataset of violations. Note that the p-values were not adjusted since they are too small and the adjustment would have no impact.

Considering the numbers presented in \autoref{tab:rq1} other than the proportions of fully-repaired violations, we noticed that \CSIDEA and \Styler are the most reliable tools in the sense of delivering to an end-user either a repaired source code or, in the worst-case scenario, the code with the same violation. It is not the same case of \Naturalize and \CodeBuff, which had higher rates of delivering broken source code. They were, however, designed for a different goal and do not take into account the Checkstyle ruleset of the project like \Styler and \CSIDEA do. Yet, they are relevant for our experiment since they are the state of the art of machine-learning-based code formatters. Our results show the need for specialized, focused tools to repair Checkstyle violations.

\begin{table}[b]
    \caption{Repairability results on the \nbViolationsEvaluationToPrint real violations per tool.}
    \label{tab:rq1}
    \centering
    \begin{tabular}{@{}p{.13\textwidth} | p{.136\textwidth} p{.15\textwidth} | p{.155\textwidth} p{.125\textwidth} p{.125\textwidth}@{}}
        \toprule
        {} & \multicolumn{2}{c|}{Repaired violations} & \multicolumn{3}{c}{Not repaired violations} \\
        Tool & \multicolumn{1}{c}{\scriptsize{fully repaired}} & \multicolumn{1}{c|}{\scriptsize{new violations}} & \multicolumn{1}{c}{\scriptsize{same violation}} & \multicolumn{1}{c}{\scriptsize{same + new}} & \multicolumn{1}{c}{\scriptsize{broken}} \\
        \midrule
        
        \Styler & \ChartWithPercentageOnly{11008}{\nbViolationsEvaluation} & \ChartWithPercentageOnly{3121}{\nbViolationsEvaluation} & \ChartWithPercentageOnly{11965}{\nbViolationsEvaluation} & \ChartWithPercentageOnly{117}{\nbViolationsEvaluation} & \ChartWithPercentageOnly{580}{\nbViolationsEvaluation} \\
        
        \CSIDEA & \ChartWithPercentageOnly{13376}{\nbViolationsEvaluation} & \ChartWithPercentageOnly{2610}{\nbViolationsEvaluation} & \ChartWithPercentageOnly{3179}{\nbViolationsEvaluation} & \ChartWithPercentageOnly{7626}{\nbViolationsEvaluation} & \ChartWithPercentageOnly{0}{\nbViolationsEvaluation} \\
        
        \Naturalize & \ChartWithPercentageOnly{3900}{\nbViolationsEvaluation} & \ChartWithPercentageOnly{4132}{\nbViolationsEvaluation} & \ChartWithPercentageOnly{7395}{\nbViolationsEvaluation} & \ChartWithPercentageOnly{2225}{\nbViolationsEvaluation} & \ChartWithPercentageOnly{9139}{\nbViolationsEvaluation} \\
        
        \CodeBuff & \ChartWithPercentageOnly{5326}{\nbViolationsEvaluation} & \ChartWithPercentageOnly{4463}{\nbViolationsEvaluation} & \ChartWithPercentageOnly{2642}{\nbViolationsEvaluation} & \ChartWithPercentageOnly{3309}{\nbViolationsEvaluation} & \ChartWithPercentageOnly{11051}{\nbViolationsEvaluation} \\
        
        \bottomrule
    \end{tabular}
\end{table}

\begin{table}[t]
    \caption{Contingency tables of the repairability of \Styler versus the other tools.}
    \label{tab:rq1-contingency-tables}
    \centering
    \begin{tabular}{@{}l l  r r  c@{}}
        \toprule
        {} & {} & \multicolumn{2}{c}{\Styler} & McNemar test \\
        {} & {} & repaired & not repaired & p-value \\
        \midrule
        
        \multirow{2}{*}{\CSIDEA} & repaired & \numprint{9039} & \numprint{4337} & \multirow{2}{*}{$<$0.00001} \\
        {} & not repaired & \numprint{1969} & \numprint{11446} &  \\[0.2cm]
        
        \multirow{2}{*}{\Naturalize} & repaired & \numprint{3290} & \numprint{610} & \multirow{2}{*}{$<$0.00001} \\
        {} & not repaired & \numprint{7718} & \numprint{15173} &  \\[0.2cm]
        
        \multirow{2}{*}{\CodeBuff} & repaired & \numprint{3530} & \numprint{1796} & \multirow{2}{*}{$<$0.00001} \\
        {} & not repaired & \numprint{7478} & \numprint{13987} &  \\

        \bottomrule
    \end{tabular}
\end{table}

In addition, we observed that some violation types, i.e., violations of different Checkstyle rules, occur in a much higher frequency than others in our dataset (see \autoref{tab:real-violation-dataset}). This might cause bias in the results presented in \autoref{tab:rq1}. Because of that, we performed a normalization of the data by sub-sampling the most frequent violation types. In this way, we obtained a sub-sample of violations that contains the same number of instances for all violation types. We ignored the less frequent ones to avoid using too few instances. For that, we calculated the median of the distribution of the violation types, which is 274, and used it as the minimum number of instances for including Checkstyle rules in the analysis. Then, we randomly selected 274 violations of the included rules. In the end, the analysis comprises half of the rules (13) and \numprint{3562} violations. The normalized results are presented in \autoref{tab:rq1-normalized}. All the tools are impacted positively in terms of fully-repaired violations. However, we note that the normalized results present a different ranking of the tools' performance, where \Styler outperforms \CSIDEA. \CSIDEA is the tool most negatively impacted by the normalization because it increases only 4\% fully-repairs, while the other three tools considerably increase their repairability by 7\%--20\%. This suggests that \CSIDEA performs better than the other tools on violation types that are frequent in our dataset, which is investigated in more detail in the next section, for answering RQ \#2. Finally, we also performed McNemar test in the normalized results, as shown in \autoref{tab:rq1-contingency-tables-normalized}. Considering $\alpha = 0.05$, the results show that \Styler and any other tool have a statistically significant different proportion of errors on the sub-sample too.

\begin{table}[t]
    \caption{Normalized repairability results based on \numprint{3562} violations equally distributed across 13 violation types.}
    \label{tab:rq1-normalized}
    \centering
    \begin{tabular}{@{}p{.13\textwidth} | p{.136\textwidth} p{.15\textwidth} | p{.155\textwidth} p{.125\textwidth} p{.125\textwidth}@{}}
        \toprule
        {} & \multicolumn{2}{c|}{Repaired violations} & \multicolumn{3}{c}{Not repaired violations} \\
        Tool & \multicolumn{1}{c}{\scriptsize{fully repaired}} & \multicolumn{1}{c|}{\scriptsize{new violations}} & \multicolumn{1}{c}{\scriptsize{same violation}} & \multicolumn{1}{c}{\scriptsize{same + new}} & \multicolumn{1}{c}{\scriptsize{broken}} \\
        \midrule
        
        \Styler & \ChartWithPercentageOnly{2189}{3562} & \ChartWithPercentageOnly{440}{3562} & \ChartWithPercentageOnly{843}{3562} & \ChartWithPercentageOnly{47}{3562} & \ChartWithPercentageOnly{43}{3562} \\
        
        \CSIDEA & \ChartWithPercentageOnly{1936}{3562} & \ChartWithPercentageOnly{793}{3562} & \ChartWithPercentageOnly{509}{3562} & \ChartWithPercentageOnly{324}{3562} & \ChartWithPercentageOnly{0}{3562} \\
        
        \Naturalize & \ChartWithPercentageOnly{791}{3562} & \ChartWithPercentageOnly{795}{3562} & \ChartWithPercentageOnly{580}{3562} & \ChartWithPercentageOnly{249}{3562} & \ChartWithPercentageOnly{1147}{3562} \\
        
        \CodeBuff & \ChartWithPercentageOnly{1221}{3562} & \ChartWithPercentageOnly{1023}{3562} & \ChartWithPercentageOnly{165}{3562} & \ChartWithPercentageOnly{348}{3562} & \ChartWithPercentageOnly{805}{3562} \\
        
        \bottomrule
    \end{tabular}
\end{table}

\begin{table}[t]
    \caption{Contingency tables of the repairability of \Styler versus the other tools considering the normalized results.}
    \label{tab:rq1-contingency-tables-normalized}
    \centering
    \begin{tabular}{@{}l l  r r  c@{}}
        \toprule
        {} & {} & \multicolumn{2}{c}{\Styler} & McNemar test \\
        {} & {} & repaired & not repaired & p-value \\
        \midrule
        
        \multirow{2}{*}{\CSIDEA} & repaired & \numprint{1457} & \numprint{479} & \multirow{2}{*}{$<$0.00001} \\
        {} & not repaired & \numprint{732} & \numprint{894} &  \\[0.2cm]
        
        \multirow{2}{*}{\Naturalize} & repaired & \numprint{736} & \numprint{55} & \multirow{2}{*}{$<$0.00001} \\
        {} & not repaired & \numprint{1453} & \numprint{1318} &  \\[0.2cm]
        
        \multirow{2}{*}{\CodeBuff} & repaired & \numprint{937} & \numprint{284} & \multirow{2}{*}{$<$0.00001} \\
        {} & not repaired & \numprint{1252} & \numprint{1089} &  \\

        \bottomrule
    \end{tabular}
\end{table}

\answer{
\textbf{RQ \#1: \rqone}\\
\Styler repaired 41\% (\numprint{11008}/\nbViolationsEvaluationToPrint)
of the Checkstyle formatting violations found in the wild. \Styler outperformed the machine learning systems \Naturalize and \CodeBuff. \CSIDEA outperformed \Styler on our entire dataset of violations, with a repairability of 50\%. However, this is not the case when we consider a sub-sample of the dataset by normalizing the number of instances according to violation types, which suggests that \CSIDEA performed better than the other tools on violations of frequent types in our dataset. In addition, note that \CSIDEA depends on the code formatter of the IntelliJ IDE, whereas \Styler's approach is fully automated and hence more appropriate for handling new and configurable rules.
}

\subsection{Violation-type-based repairability (RQ \#2)}

\begin{figure}[t]
    \centering
    \includegraphics[scale=0.75]{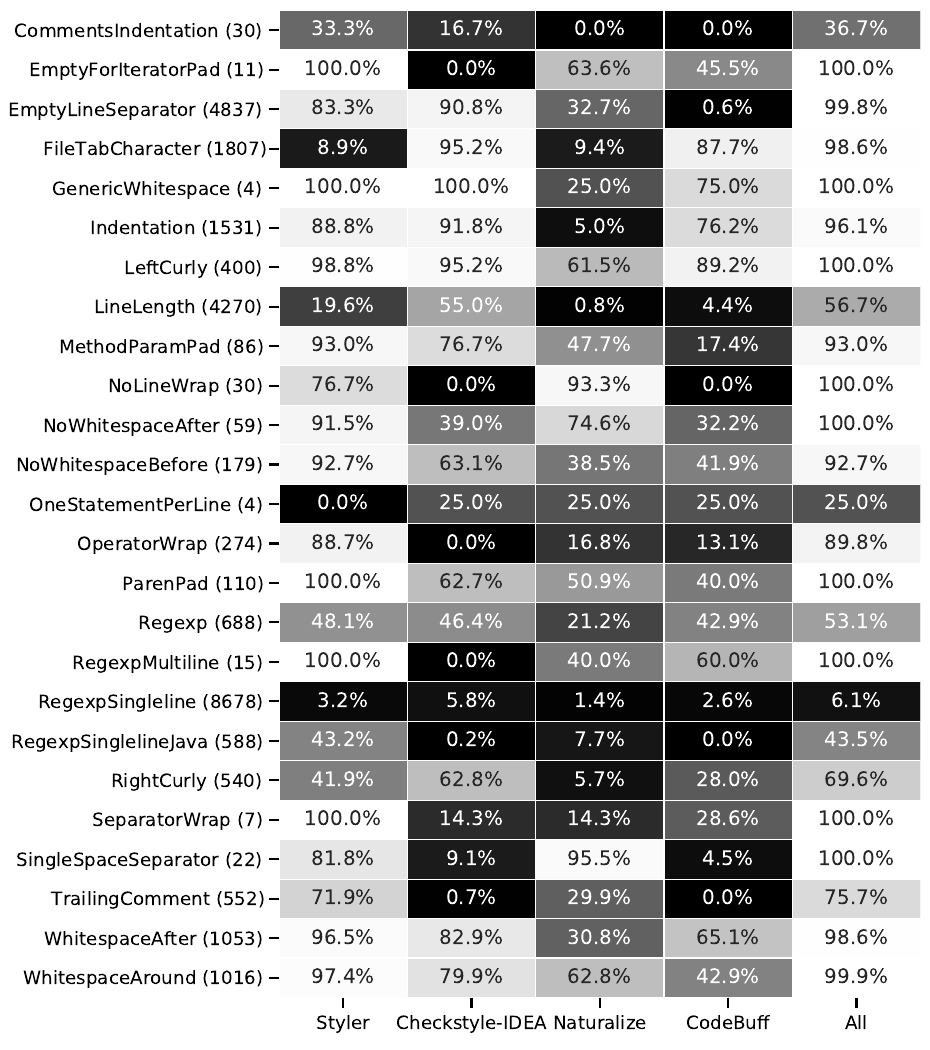}
    \caption{Types of Checkstyle violation repaired per tool.}
    \label{fig:rq2}
\end{figure}

To answer RQ \#2, we investigated the extent to which \Styler and the other three systems repair different Checkstyle violation types, i.e., violations of different Checkstyle rules. \autoref{fig:rq2} shows the Checkstyle violations fully repaired by the systems per violation type in a heatmap. The color scale is from black to white, where black represents 0\% of fully-repaired violations and white represents 100\% (i.e., the lighter, the better).

\Styler and \Naturalize repaired violations of 24/25 Checkstyle rules, which is the highest coverage of rules considering all the four tools. \CSIDEA and \CodeBuff fixed violations of 21 rules. Surprisingly, \Naturalize produced fixes for a higher number of violation types than \CSIDEA, even though it does not consider the Checkstyle ruleset of projects because of its different goals. \CodeBuff performed relatively well considering that it does not target Checkstyle violations as \Naturalize. These facts suggest that our idea of employing a machine learning approach for repairing format checker violations is promising.

The reason for the high overall repairability of \CSIDEA (found in RQ \#1) is that it outperformed the other tools in the five most frequent rules in our dataset: \emph{RegexpSingleline}, \emph{EmptyLineSeparator}, \emph{LineLength}, \emph{FileTabCharacter}, and \emph{Indentation}.
This explains why \CSIDEA is the leading tool in terms of repairability on the entire dataset but not on a sample of it, as shown in RQ \#1.
\Styler, on the other hand, had a perfect success rate in rules that are not that frequent: \emph{EmptyForIteratorPad}, \emph{GenericWhitespace}, \emph{ParenPad}, \emph{RegexpMultiline}, and \emph{SeparatorWrap}.
Interestingly, \CSIDEA did not repair, at least not fully, any violation of two of these rules. This indicates that the tools are complementary to each other.
Moreover, \Styler performed very well, with at least 80\% of repaired violations, in the rules that are related to horizontal whitespace between two Java tokens, such as \emph{MethodParamPad}, \emph{NoWhitespaceAfter}, and \emph{WhitespaceAround}.
For developers, even if fixing these types of violations is easy, they may have dozens of them, which could be overwhelming. To that extent, automation is still valuable.
Moreover, \Styler is able to repair these violation types for which one would not need to put engineering effort to write the repair code.
Finally, we observed that all tools performed poorly on violations of the most frequent type in our dataset, i.e., \emph{RegexpSingleline}.

\answer{
\textbf{RQ \#2: \rqtwo}\\
\Styler and \Naturalize repaired violations of a greater diversity of Checkstyle rules (24/25) than the other tools (\CSIDEA and \CodeBuff: 21).
\Styler performed well for fixing violations related to horizontal whitespace between Java tokens.
For some rules, \Styler fixed all violations while \CSIDEA did not fix any and, for other rules, \CSIDEA had a much higher repairability than \Styler, suggesting that they can be considered as complementary in practice. Finally, we confirmed that \CSIDEA outperformed the other tools on the five most frequent violation types in our dataset.
}

\subsection{Unsuccessful repair cases (RQ \#3)}

\Styler repaired violations of 24/25 Checkstyle rules, but it did not perform well for some rules as shown in RQ \#2. To understand in which cases \Styler does not successfully generate repairs, we manually analyzed violations of the Checkstyle rules for which \Styler repaired less than 50\% violations. The analysis was ad-hoc, where, for each rule, both repaired and non-repaired violations were investigated so that patterns of non-repaired violations or their contexts could be identified. We present the cases of unsuccessful repair we found as follows.

\Styler encodes a violation according to the source code position returned by a format checker which is, in this case, Checkstyle. However, in some cases, this position is not where a fix should be applied. For instance, for a violation of the type \emph{OneStatementPerLine}\footnote{\url{https://kth.github.io/styler-experiments/\#!/violation/Angel-ML-angel/339}}, a line break should be added in the column 19 or 20 of the line 42, just after the first statement. However, Checkstyle returns column 31, which is the end of the second statement. In such a case, \Styler tried to repair the violation in an inappropriate location.

Several non-repaired violations were inside comments. For instance, we found lines of comments exceeding the maximum length of characters, therefore triggering violations of the type \emph{LineLength}\footnote{\url{https://kth.github.io/styler-experiments/\#!/violation/opencb-opencga/173}}. We also found tab characters inside comments, triggering \emph{FileTabCharacter} violations\footnote{\url{https://kth.github.io/styler-experiments/\#!/violation/actiontech-txle/13}}. These violations also happen with strings\footnote{\url{https://kth.github.io/styler-experiments/\#!/violation/apache-iotdb/463}}. \Styler does not handle cases in which comments or strings should be modified. This is a limitation of \Styler due to its tokenization. Comments and strings are tokenized as a single token, i.e., \Styler does not take into account the separation of words.

Moreover, we found several occurrences of a case in which \Styler repaired a given violation but then another existing one, which was not previously reported by Checkstyle, was triggered. This case of only one violation being reported when multiple ones exist in files is recurrent and happens with \emph{FileTabCharacter} violations. We were not aware of that at the time we built the real violation dataset, but when a file contains more than one tab character, Checkstyle reports only the first instance of it. In some cases, \Styler repaired the first instance\footnote{\url{https://kth.github.io/styler-experiments/\#!/violation/actiontech-txle/19}}, but the next one was then raised by Checkstyle. Even though \Styler repaired the originally reported violation, it was not counted as a repaired violation in our study. Note that one criterion to select files containing violations when building the dataset was the existence of a single violation in them (see \autoref{sec:data-collection}). This was a decision we made to guarantee we could automatically check if a given violation was fixed. In such a case with \emph{FileTabCharacter} violations, however, we could not check that precisely.

Finally, we observed that \emph{RegexpSingleline} violations are the most frequent ones in our dataset and are poorly handled not only by \Styler, but all tools (see the last column of \autoref{fig:rq2}). When analyzing the violations related to this rule and other regex ones, we found out that many violations are not related to formatting. Some examples are violations related to missing, wrong, or duplicated license header\footnote{\url{https://kth.github.io/styler-experiments/\#!/violation/neo4j-neo4j/0}} and the usage of specific patterns, such as a tag in javadoc, that are forbidden in some projects\footnote{\url{https://kth.github.io/styler-experiments/\#!/violation/apache-usergrid/0}}. Since these violations are not about formatting, they are not in the targeted violation types of \Styler and other tools. However, the occurrence of these violations is very frequent in our dataset and, consequently, the repairability of the tools for such regex violations is impacted. For instance, our dataset contains \numprint{8678} \emph{RegexpSingleline} violations, and \numprint{8102} (93\%) of them are non-formatting violations. The overall repairability results about that rule, as presented in \autoref{fig:rq2}, are 3.2\% for \Styler, 5.8\% for \CSIDEA, 1.4\% for \Naturalize, and 2.6\% for \CodeBuff. Adjusting the repairability results of the tools by considering only the 576 \emph{RegexpSingleline} violations that are about formatting, we found out that \Styler, \CSIDEA, \Naturalize, and \CodeBuff repaired 45.1\%, 86.8\%, 21\%, 38.9\% of \emph{RegexpSingleline} violations, respectively.

\answer{
\textbf{RQ \#3: \rqthree}\\
\Styler does not generate a correct repair when Checkstyle returns a source code position other than the one to be modified and when the violation is inside comments or strings. Moreover, \Styler produces repairs that make Checkstyle raise non-originally-reported violations. These cases in which \Styler failed to generate a correct repair relate to some limitations of \Styler, which can be further addressed in new studies. Finally, we found out that most of the violations of regex rules are not about formatting, which explains why the repairability of the tools is low for these rules.
}

\subsection{Size of the repairs (RQ \#4)}

One aspect of repair quality is the size of the diff between the source code with a formatting violation and the repaired source code. There might be different repairs for the same violation that pass all Checkstyle rules, but the one with the smaller diff size would be preferable for being the least disrupting for the developers. In the context of a pull request on GitHub, a smaller diff is usually considered as easier to review and merge \citep{smallpr}.

\begin{figure}[t]
    \centering
    \includegraphics[scale=0.62]{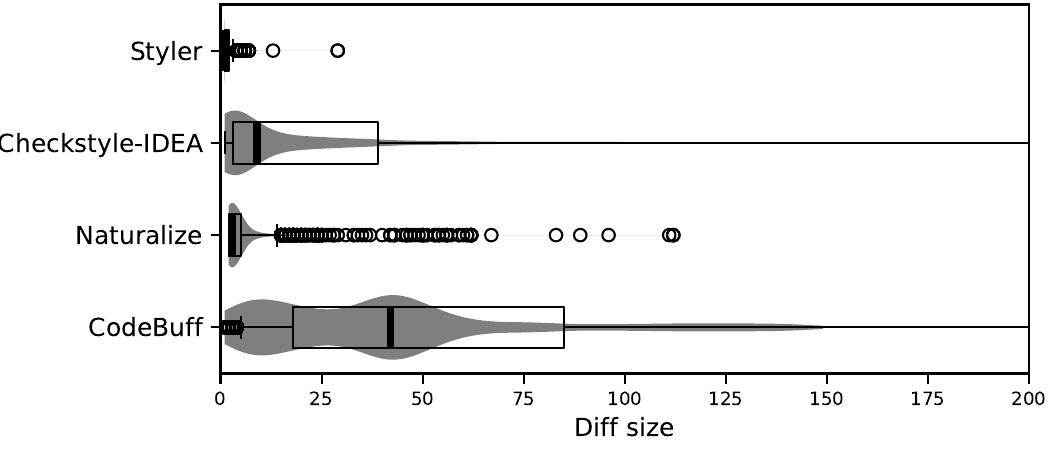}
    \caption{Size of the repairs per tool. The two boxplot whiskers represent the 5th and the 95th percentiles.}
    \label{fig:repair-size}
\end{figure}

To answer RQ \#4, we calculated the diff size, in number of lines, of the repairs generated by \Styler, \CSIDEA, \Naturalize, and \CodeBuff.
\autoref{fig:repair-size} shows the distributions of diff size per tool.
We observed that the distributions of the repairs generated by \Styler and \Naturalize have the smallest medians, which are equal to one and three changed lines, respectively. Yet, they suffer from a few bad cases (the right-hand part of the distributions), mainly \Naturalize.
\CSIDEA and \CodeBuff produced larger repairs, with medians equal to nine and 42, respectively. In the worst cases, they produced several repairs with more than 200 changed lines, which can be seen by the fact that their 95th percentiles are not shown in \autoref{fig:repair-size}. On the other hand, the 95th percentile of \Styler is three.
We performed Wilcoxon rank-sum test to verify if the distributions of diff sizes obtained by \Styler and the other tools are significantly different from one another. We found \textit{p-value} $<$ 0.00001 when testing \Styler with all the other tools. Considering $\alpha=0.05$, we rejected the null hypothesis, which means that the distribution of \Styler is significantly different from the other ones.

\answer{
\textbf{RQ \#4: \rqfour}\\
The size of the repairs produced by \Styler is usually small. \Styler had the smallest median repair size of one changed line, followed by \Naturalize, with a median size of three lines.
\CSIDEA and \CodeBuff produced larger repairs.
The ability to produce small diffs is an important property for code review and pull-request-based development, hence our results show that \Styler can be realistically used in a modern software development context.
}

\subsection{Performance (RQ \#5)}

To investigate if \Styler can be used in practice, we measured the time \Styler spent on the real violation dataset.
\autoref{tab:rq4} shows the minimum, median, average, and maximum time spent on the \nbProjectsEvaluationToPrint projects, split over the different steps of the \Styler workflow.
For training data generation, \Styler took at least 15 minutes and up to six hours, which is the maximum execution time allowed by our experimental setup (see \autoref{sec:execution}). The median time for training data generation was 45 minutes. To tokenize the training data, \Styler took around two minutes on average, and a maximum of 14 minutes for training the models.
The entire training process of \Styler (data generation + tokenization + model training) took around one hour and a half on average.
This can be considered just fine since the training is meant to happen only when the coding conventions used in a project change (i.e., the Checkstyle ruleset file).
After \Styler is trained for a given project, it takes on average two seconds to predict a repair, which is fast enough to be used in IDEs or in continuous integration environments.

\begin{table}[ht]
    \caption{Statistics on the performance of \Styler.}
    \label{tab:rq4}
    \centering
    \begin{threeparttable}[b]
        \begin{tabular}{@{}l|c c c|c@{}}
            \toprule
            {} & \multicolumn{3}{c|}{\textit{Training}} & \textit{Prediction} \\
            {} & Data generation & Tokenization & Models & Average Time \\ 
            Step (\autoref{fig:styler})\tnote{a}~: & A & B & C & E\hspace*{0.6pt}$\rightarrow$\hspace*{0.6pt}I \\
            \midrule
            Min & 00:15:58 & 00:00:07 & 00:04:33 & 01.380 s/err \\
            Med & 00:45:19 & 00:01:36 & 00:07:07 & 02.133 s/err \\
            Avg & 01:23:05 & 00:01:53 & 00:07:13 & 02.267 s/err \\
            Max & 06:12:33 & 00:05:38 & 00:14:01 & 05.723 s/err \\
            \bottomrule
        \end{tabular}
        \begin{tablenotes}
            \item[a]{The steps were executed in a computer containing a processor Intel(R) Core(TM) i9-10980XE CPU @ 3.00GHz and 125GiB system memory. For training the models, we used GPUs GeForce RTX 2080 Ti.}
        \end{tablenotes}
    \end{threeparttable}
\end{table}

\answer{
\textbf{RQ \#5: \rqfive}\\
On average, \Styler needs about one hour and a half for training and two seconds for predicting a repair.
The training time is not an issue since it only happens when the Checkstyle ruleset file of a project changes.
The prediction time relates to usability: our results show that \Styler can be used in IDEs or CI in a practical setting.
}

\subsection{Technical analysis on \Styler (RQ \#6)}

At prediction time, \Styler used two trained LSTM models, each one based on a different training data generation protocol: \protocolOne and \protocolTwo. In RQ \#6, we investigated how the two protocols contribute to the final output of \Styler. We found out that \Styler fixed 967 violations exclusively with the \protocolOne-based model and \numprint{2581} violations exclusively with the \protocolTwo-based model. \numprint{7460} violations were fixed with both models. This shows that the model based on the \protocolTwo protocol is more effective than the model based on the \protocolOne one. In a real case scenario, one could consider using \Styler only with the \protocolTwo-based model and still obtain fixes for 91\% of what \Styler can repair. This would reduce the time for training \Styler.

\answer{
\textbf{RQ \#6: \rqsix}\\
The LSTM model based on the \protocolTwo protocol is more effective than the model based on the \protocolOne protocol. Even though the latter exclusively contributes to the overall repairability of \Styler with 9\% of the fixes, in a real case scenario, one could use only the former to reduce the time for training \Styler.
}

\section{Discussion}\label{sec:discussion}

We discuss, in this section, machine learning versus rule-based approaches, as well as the threats to the validity of our study.

\subsection{Machine learning versus rule-based approaches}

\Styler employs a machine-learning-based approach for repairing formatting convention violations.
An alternative approach would be a rule-based one. In such a case, there would be at least one transformation to be applied in the code per format checker rule. However, the engineering of a transformation for every single linter rule is time-consuming. While this is costly, this might be even impractical for highly configurable linters such as Checkstyle, because the rule-based repair system would need to have different transformations for the same linter rule due to the configurable properties.
On the contrary, a machine learning approach does not require costly human engineering. It is able to infer transformations for a diverse set of linter rules. Our experiments have validated this property in the context of formatting violations raised by Checkstyle.
However, since our approach is far from being perfect and does not work well for certain rules, one avenue for future research is the development of a rule-based system for simple, non-highly configurable rules, to be complementary to \Styler.

\subsection{Threats to validity}

\Styler generates training data for learning how to repair violations based on the Checkstyle ruleset file contained in a given project. This means that \Styler assumes that all formatting rules contained in such a file are valid. In practice, however, developers might ignore the violations of certain rules. Our experiment does not take this scenario into account, thus we do not claim that all the fixes produced by \Styler are necessarily relevant for developers.

The real violation dataset contains Checkstyle violations mined from GitHub repositories. It is to be noted that it does not cover all existing Checkstyle formatting rules.
Moreover, the dataset might not be representative of the real distribution of the 25 rules in the real world. Consequently, future research is needed to strengthen the validity of our study.

At the end of the violation collection process for building the dataset, we removed duplicate Java files according to file contents. However, there might still exist duplicate violations in our dataset. A file containing a violation might have changed, but that change might not be related to the line where the violation exists. Therefore, the same violation would exist in both versions of the file, but since the overall file contents are different, they are both kept in our dataset, which might create noise in it.

Another threat related to the creation of the dataset is that, when selecting violating files, we chose only the ones containing a single Checkstyle violation. We performed this selection so that we could accurately check if the violation was correctly repaired by the tools. Files containing more than one violation would make it hard to automatically check the correctness of repairs because once a violation is repaired, the location of the other ones in the file could be different. Therefore, our results are based on single-violation files, and future investigations on multiple-violation files are needed.

Finally, to compare the quality of the repairs produced by \Styler with the repairs produced by the other three tools, we measured the size in lines of the diff between the original and repaired program versions. However, the diff size is only one dimension for comparing the tools, which only approximates the developer's perception of formatting repairs. User studies, such as proposing formatting repairs to developers, are interesting future experiments to further investigate the practical value of this research.

\section{Related work}\label{sec:related-work}

\Styler aims to repair formatting violations raised by linters. Linters are a kind of automated static analysis tool. In this section, we first present works on the usage of static analysis tools. Then, we present systems that share similar goals with \Styler, which are systems that target linter violation repair and systems that fix source code formatting. Finally, we present works related to \Styler in terms of used technique, i.e., machine learning, for repairing compiler errors and behavioral bugs.

\subsection{The usage of automated static analysis tools}

Static analysis tools have been the subject of investigation in recent research.
\cite{Zampetti2017} investigated their usage in 20 popular Java open source projects hosted on GitHub that use Travis CI to support CI activities.
They first found out that the projects use seven static analysis tools---Checkstyle, FindBugs \citep{Ayewah2008findbugs}, PMD\footnote{\url{https://pmd.github.io/}, last access: 2020-07-13}, License Gradle Plugin\footnote{\url{https://github.com/hierynomus/license-gradle-plugin}, last access: 2020-07-13}, Apache Rat\footnote{\url{https://creadur.apache.org/rat/}, last access: 2020-07-13}, Clirr\footnote{\url{http://www.mojohaus.org/clirr-maven-plugin/}, last access: 2020-07-13}, and jDepend\footnote{\url{http://www.mojohaus.org/jdepend-maven-plugin/}, last access: 2020-07-13}---being Checkstyle the most used one.
About the integration of static analysis tools in CI pipelines, they found out that build breakages due to those tools are mainly related to adherence to coding conventions, while breakages related to likely bugs or vulnerabilities occur less frequently. \cite{Zampetti2017} discuss that some tools are sometimes configured to just produce warnings without breaking the build, possibly because of the high number of false positives.

\cite{Vassallo2018} investigated the usage of static analysis tools from the perspective of the development context in which these tools are used. For that, they surveyed 42 developers and interviewed 11 industrial experts that integrate static analysis tools in their workflow.
They found out that static analysis tools are used in three main development contexts, which are local environment, code review, and continuous integration. Moreover, they also found out that developers consider different warning types depending on the context, e.g., when performing code review they mainly look at style conventions and code redundancies.

\cite{Marcilio2019} focused on one specific static analysis tool: SonarQube\footnote{\url{https://www.sonarqube.org/}, last access: 2020-07-13}. Through an online survey with 18 developers from different organizations, they found out that most respondents agree that the issues reported by static analysis tools are relevant for improving the design and implementation of software.

\subsection{Linter violation repair and code formatters}

\textit{Linter violation repair.}
There are some tools to fix violations of rules checked by linters.
Considering academic systems, there are Phoenix \citep{bavishi2019phoenix}, which repairs violations of rules checked by FindBugs \citep{Ayewah2008findbugs}, and Getafix \citep{Bader2019getafix}, which focuses on rules checked by Infer \citep{Calcagno2015infer} and Error Prone \citep{Aftandilian2012errorprone}. These tools learn fix patterns by mining past human-written fixes for linter violations. Another tool is SpongeBugs \citep{marcilio2020spongebugs}, which repairs violations of rules checked by the two well-known static analyzers SonarJava and SpotBugs with fixed repair templates.
\Styler shares with these tools the goal of generating patches for linter violations. However, while the mentioned tools focus on rules related to bugs and code smells, \Styler focuses on formatting.
In addition, there is C-3PR \citep{carvalho2020c}, which does not generate patches itself but proposes fixes through pull request on GitHub generated by linter violation repair tools.

Beyond those academic systems, there are other tools that repair violations found by linters. Related to formatting rules, there is, for instance, ESLint, which is a linter for JavaScript, and it also includes automated solutions to repair violations raised by it.

\vspace{5pt}
\noindent\textit{Code formatters.}
A way to enforce formatting conventions lies in code formatters (also known as pretty-printers).
In \autoref{sec:systems-to-compare}, we described \Naturalize \citep{Allamanis2014naturalize} and \CodeBuff \citep{Parr2016codebuff}. \Naturalize recommends fixes for coding conventions related to naming and formatting in Java programs, and \CodeBuff infers formatting rules to any language given a grammar.
Similar to the idea behind \CodeBuff, \cite{Reiss2007} had previously experimented with different learning algorithms and feature set variations to learn the style of a given corpus so that it could be applied to arbitrary code.
More recently, \cite{Markovtsev2019} presented \textsc{Style-Analyzer}, which helps developers to fix code formatting during code reviews. \textsc{Style-Analyzer} mines the formatting style of the git repository under analysis and expresses the found format patterns with compact human-readable rules. Then, it suggests style inconsistency fixes in the form of code review comments.

Beyond those academic systems, there are code formatters such as google-java-format\footnote{\url{https://github.com/google/google-java-format/}, last access: 2020-07-13}, which reformats source code according to the Google Java Style Guide\footnote{\url{http://checkstyle.sourceforge.net/reports/google-java-style-20170228.html}, last access: 2020-07-13}. However, these formatters are usually not configurable or require manual tweaking, which is a tedious process for developers. This is a problem because not all developers are ready to follow a unique convention style. \Styler, on the other hand, is generic and automatically captures the conventions used in a project to fix formatting violations.

Finally, there is the \CheckStyleIDEA plugin for IntelliJ \citep{checkstyle-idea}, which we used to compare \Styler with. \CheckStyleIDEA provides both real-time and on-demand scanning of Java files with Checkstyle from within IDEA. It also uses the Checkstyle ruleset of projects to configure the formatter available in IntelliJ, making it possible to repair Checkstyle formatting violations. However, it is limited in repairing violations of a great number of Checkstyle rules as shown in RQ \#2 and creates large repairs as shown in RQ \#4.

\subsection{Learning for repairing compiler errors and behavioral bugs}

\noindent\textit{Learning for repairing compiler errors.}
There are related works in the area of automatic repair of compiler errors. In this case, the compiler syntax rules are the equivalent of the formatting rules.  
There, recurrent neural networks and token abstraction have been used to fix syntactic errors \citep{Bhatia2018}. 
In DeepFix, \cite{Gupta2017} use a language model for repairing syntactic compilation errors in C programs. Out of 6,971 erroneous C programs, DeepFix was able to completely repair 27\% and partially repair 19\% of the programs.
Later, \cite{Ahmed2018} proposed TRACER, which outperformed DeepFix, repairing 44\% of the programs.
\cite{Santos2018} confirmed the efficiency of LSTM over n-grams and of token abstraction for single token compiling errors.
These approaches do not target formatting violations, which is the target of \Styler.

\vspace{5pt}
\noindent\textit{Learning for repairing behavioral bugs.}
As for repairing compiler errors, there are also learning systems for repairing behavioral bugs, those that, for instance, break test cases.
\cite{Tufano2018} investigated the feasibility of using Neural Machine Translation techniques for learning bug-fixing patches for real defects. They mined millions of buggy and patched program versions from the history of GitHub repositories and abstracted them to train an Encoder-Decoder model. The model was able to fix hundreds of unique buggy methods in the wild.
\cite{Chen2019} proposed SequenceR, a program repair tool based on sequence-to-sequence learning focused on one-line fixes. In an experiment with Defects4J \citep{Just2014}, SequenceR was shown to be able to learn how to repair behavioral bugs by generating patches that pass all tests. \Styler and SequenceR share the same idea for formatting violation and bug encoding.

\section{Conclusion}\label{sec:conclusion}

In this paper, we presented \Styler, which implements a novel approach to repair formatting violations raised by format checkers. \Styler creates a corpus of violations, learns from it, and predicts fixes for new violations, using machine learning. Currently, its implementation supports Checkstyle, a popular linter for Java programs.
Our experimental results on \nbViolationsEvaluationToPrint real Checkstyle violations showed that \Styler repairs real violations from a diverse set of Checkstyle rules and performs better for fixing violations related to horizontal whitespace between Java tokens than for fixing violations related to tabulations and line length.
Moreover, \Styler produces smaller repairs than the compared systems, and its prediction time is low, which suggests that it can be used in development environments such as IDEs. Finally, we identified cases in which \Styler does not succeed to generate correct repairs, e.g., for Checkstyle violations inside comments or strings. These findings can guide improvements in \Styler and help researchers and developers to understand \Styler's limitations.

There are several interesting avenues for future research.
First, improvements on the violation injection protocols for creating training data can be done to improve the representativeness of seeded formatting violations. This might increase the repairability of \Styler.
Second, user studies can be conducted, where repairs predicted by \Styler are proposed to developers through, for instance, pull requests on GitHub. This type of study would bring practical insights on the potential of \Styler.
Third, \Styler could be integrated into development environments, such as IDEs and social coding sites, for supporting the mentioned user studies and possibly for developers to use \Styler.
Fourth, other linters could be plugged in \Styler so it could be applicable on projects that use linters other than Checkstyle.
Fifth, since \Styler does not work well for certain rules, the development of a rule-based system for simple, non-highly configurable rules, could be beneficial to complement \Styler.
Finally, the overall idea behind \Styler could be tried out to repair other linter violations beyond purely formatting ones.

\section*{Acknowledgments}

We thank the anonymous reviewers for their valuable feedback that made us improve this paper. We also thank Terence Parr, Miltos Allamanis, Vadim Markovtsev, Hugo Mougard, Matias Martinez, and the ASSERT members who gave feedback on a draft version of this paper. Fernanda Madeiral would like to thank Thomas Durieux for his support on this work.

This work was partially supported by the Swedish Foundation for Strategic Research under the TrustFull project, and by the Knut and Alice Wallenberg Foundation under the Wallenberg AI, Autonomous Systems, and Software Program (WASP). Some experiments were performed on resources provided by the Swedish National Infrastructure for Computing.

\bibliographystyle{spbasic}
\bibliography{references}

\end{document}